\documentclass[aps,
               prx,
               showpacs,
               amssymb,
               superscriptaddress,
               twocolumn,
              nofootinbib,
              longbibliography,
              10pt,
              floatfix]{revtex4-2}
\pdfoutput=1
\usepackage[utf8]{inputenc}
\usepackage[english]{babel}
\usepackage[T1]{fontenc}
\usepackage{amsfonts}
\usepackage{amsmath}
\usepackage{amsthm}
\usepackage{amssymb}
\usepackage[dvipsnames]{xcolor}
\usepackage[pdftex,colorlinks=true,urlcolor= blue,linkcolor=Blue,citecolor=RedViolet]{hyperref}
\usepackage{bm}
\usepackage{tikz}
\usepackage{cancel}
\usepackage{graphicx}
\usepackage{booktabs} 
\usepackage{siunitx}  
\usepackage{orcidlink}

\usepackage{lipsum}
\newtheorem{theorem}{Theorem}
\theoremstyle{definition}
\newtheorem{definition}{Definition}

\newtheorem{lemma}{Lemma}
\newtheorem{obs}{Observation}

\newcommand{\ket}[1]{\ensuremath{\left|\right.\!{#1}\!\left.\right\rangle}}

\newcommand{\bra}[1]{\ensuremath{\left\langle\right.\!{#1}\!\left.\right|}}

\newcommand{\ketbra}[2]{\ensuremath{|{#1}\rangle\!\langle{#2}|}}

\newcommand{\trho}{\tilde{\rho}}
\newcommand{\tp}{\tilde{p}}
\newcommand{\id}{\mathbb{I}}
\newcommand{\tr}[1]{\textnormal{Tr}\left\{#1\right\}}
\newcommand{\mst}[1]{\eta\Poi(#1)}
\newcommand{\ax}[1]{a(#1)}

\newcommand{\Sysu}{\ensuremath{{\hspace{-0.5pt}\protect\raisebox{0pt}{\tiny{$S$}}}}}
\newcommand{\Sys}{\ensuremath{_{\hspace{-0.5pt}\protect\raisebox{0pt}{\tiny{$S$}}}}}
\newcommand{\Poi}{\ensuremath{_{\hspace{-0.5pt}\protect\raisebox{0pt}{\tiny{$P$}}}}}
\newcommand{\SP}{\ensuremath{_{\hspace{-0.5pt}\protect\raisebox{0pt}{\tiny{$S\hspace*{-1pt}P$}}}}}
\newcommand{\SPi}{\ensuremath{_{\hspace{-0.5pt}\protect\raisebox{0pt}{\tiny{$S:P_1,P_2$}}}}}
\newcommand{\SPiN}{\ensuremath{_{\hspace{-0.5pt}\protect\raisebox{0pt}{\tiny{$S:\bm{P}$}}}}}
\newcommand{\Pii}{\ensuremath{_{\hspace{-0.5pt}\protect\raisebox{0pt}{\tiny{$P_i$}}}}}
\newcommand{\Piiu}{\ensuremath{{\hspace{-0.5pt}\protect\raisebox{0pt}{\tiny{$P_i$}}}}}
\newcommand{\Pij}{\ensuremath{_{\hspace{-0.5pt}\protect\raisebox{0pt}{\tiny{$P_j$}}}}}

\newcommand{\PiOne}{\ensuremath{_{\hspace{-0.5pt}\protect\raisebox{0pt}{\tiny{$P_1$}}}}}
\newcommand{\PiTwo}{\ensuremath{_{\hspace{-0.5pt}\protect\raisebox{0pt}{\tiny{$P_2$}}}}}

\newcommand{\PiN}{\ensuremath{_{\hspace{-0.5pt}\protect\raisebox{0pt}{\tiny{$\bm{P}$}}}}}
\newcommand{\Poilcg}{\ensuremath{_{\hspace{-0.5pt}\protect\raisebox{0pt}{\tiny{$\bm{P}_{\lcg}$}}}}}
\newcommand{\UBJB}[1]{U^{\text{#1}}\ensuremath{_{\hspace{-0.5pt}\protect\raisebox{0pt}{\tiny{$BJIB$}}}}}

\newcommand{\LocM}[2]{\Pi^{(#1)}_{#2}}
\newcommand{\dS}{d\Sys}
\newcommand{\NP}{N\Poi}
\newcommand{\NPcg}{N\Poi^{\mathrm{cg}}}

\newcommand{\Agr}[1]{\text{Agr}\left(#1\right)}
\newcommand{\Dis}[1]{\text{Dis}\left(#1\right)}

\newcommand{\Bias}[1]{\text{Bias}\left(#1\right)}
\newcommand{\Biasj}[2]{\text{Bias}_{#2}\left(#1\right)}

\newcommand{\fagr}[1]{\gamma^{(#1)}_{\bm{a}}}
\newcommand{\fdis}[1]{\delta^{(#1)}_{\bm{a}}}

\newcommand{\lcg}{{l_{\text{cg}}}}
\newcommand{\lcgu}{{l^{(x)}_{\text{cg}}}}
\newcommand{\axl}[2]{a(#1\vert#2)}

\newcommand{\hypgeo}[2]{%
  \operatorname{%
    {\vphantom{\mathnormal{F}}}_{#1}%
    \kern-\scriptspace
    \mathnormal{F}_{#2}%
  }%
}
\newcommand{\fagrl}[2]{\gamma^{(#1)}_{\bm{a},#2}}
\newcommand{\fdisl}[2]{\delta^{(#1)}_{\bm{a},#2}}

\newcommand{\Pilcg}{\ensuremath{_{\hspace{-0.5pt}\protect\raisebox{0pt}{\tiny{$P^{\lcg}_i$}}}}}
\newcommand{\Pilcgntext}{\ensuremath{{\hspace{-0.5pt}\raisebox{0pt}{$P^{\lcg}_i$}}}}

\newcommand{\HPD}{H_{\star}}

\begin{document}

\title{Thermodynamic~Constraints on the~Emergence~of~Intersubjectivity in Quantum~Systems}

\author{Alessandro Candeloro\,\texorpdfstring{\orcidlink{0000-0003-0582-4941}}{}}
\email{alessandro.candeloro@unipa.it}
\affiliation{School of Physics, Trinity College Dublin, Dublin 2, Ireland}
\affiliation{Università degli Studi di Palermo, Dipartimento di Fisica e Chimica - Emilio Segrè, via Archirafi 36, I-90123 Palermo, Italy}

\author{Tiago Debarba\,\texorpdfstring{\orcidlink{0000-0001-6411-3723}}{}}
\email{debarba@utfpr.edu.br}
\affiliation{Departamento Acad{\^ e}mico de Ci{\^ e}ncias da Natureza - Universidade Tecnol{\'o}gica Federal do Paran{\'a}, Campus Corn{\'e}lio Proc{\'o}pio - Paran{\'a} -  86300-000 - Brazil}
\affiliation{Atominstitut, Technische Universit{\"a}t Wien, Stadionallee 2, 1020 Vienna, Austria}
\author{Felix C. Binder\,\texorpdfstring{\orcidlink{0000-0003-4483-5643}}{}}
\email{felix.binder@tcd.ie}
\affiliation{School of Physics, Trinity College Dublin, Dublin 2, Ireland}
\affiliation{Trinity Quantum Alliance, Unit 16, Trinity Technology and Enterprise Centre, Pearse Street, Dublin 2, Ireland}

\begin{abstract}
    Ideal quantum measurement requires divergent thermodynamic resources. This is a consequence of the third law of thermodynamics, which prohibits the preparation of the measurement pointer in a fully erased, pure state required for the acquisition of perfect, noiseless measurement information. In this work, we investigate the consequences of finite resources in the emergence of intersubjectivity as a model for measurement processes with multiple observers. Here, intersubjectivity refers to a condition in which observers agree on the observed outcome (agreement), and their local random variables exactly reproduce the original random variable for the system observable (probability reproducibility). While agreement and reproducibility are mutually implied in the case of ideal measurement, finite thermodynamic resources constrain each of them. Starting from the third law of thermodynamics, we derive how the achievability of ideal intersubjectivity is affected by restricted thermodynamic resources. Specifically, we establish a no-go theorem concerning perfect intersubjectivity and present a deviation metric to account for the influence of limited resources. We further present attainable bounds for the agreement and bias that are exclusively dependent on the initial state of the environment. In addition, we show that either by cooling or coarse-graining, we can approximate ideal intersubjectivity even with finite resources. This work bridges quantum thermodynamics and the emergence of classicality in the form of intersubjectivity.
\end{abstract}

\maketitle

\section{Introduction}
In quantum mechanics, measurement of any system has to be understood not only as a dynamical mechanism but also as a fundamental principle: the associated probabilities are dictated by the Born Rule, and textbook projective measurement is postulated to result in a pure quantum state. This scenario has recently attracted attention due to the discrepancy between idealised projective measurements and the third law of thermodynamics~\cite{GuryanovaFriisHuber2018,debarba2024broadcasting, mohammady2023quantum, mohammady2025thermodynamic}. Specifically, while the measurement postulate of quantum mechanics suggests that measurement leaves the system in a pure state, the third law of thermodynamics dictates the inherent impossibility of realising such a pure state in practice~\cite{TarantoBakhshinezhadEtAl2023}. Thermodynamics is primarily concerned with macroscopic systems consisting of multiple constituent parts and their interactions. For this reason, thermodynamic considerations are essential for modelling measurement and the transition from quantum to classical.
\par 
The first foundational efforts to explore such a quantum-to-classical transition involved the study of decoherence~\cite{Zurek1991,Zurek2002,Joos2003, schlosshauer2004decoherence,Schlosshauer2007,schlosshauer2019quantum}. Over time, the framework of quantum Darwinism emerged from decoherence theory, shifting the discourse towards the view that measurement and the quantum-to-classical transition correspond to redundant, broadcasted information encoded in environmental macrofractions \cite{zurek2003decoherence,Zurek2009,zurek2022quantum}. These concepts have recently found a more structured framework in operational objectivity and Spectrum Broadcast Structures (SBS) \cite{horodecki2015quantum,korbicz2021roads}. This structure responds to the fact that quantum Darwinism, while ubiquitous~\cite{Brandao2015c}, does not per se imply agreement about measurement outcomes among observers due to quantum discord~\cite{Modi2012,Zwolak2013}, resulting in the introduction of Strong Quantum Darwinism \cite{le2019strong,Feller2021,Le2021}. These developments have provided a formalised approach to understanding the quantum-to-classical transition in terms of a specific state structure, where the salient features can be captured in terms of information broadcasting and almost perfect distinguishability of branch states \cite{touilBranchingStatesEmergent2024}. 
\par 
\begin{figure}[t]
    \centering
    \includegraphics[width=0.95\linewidth]{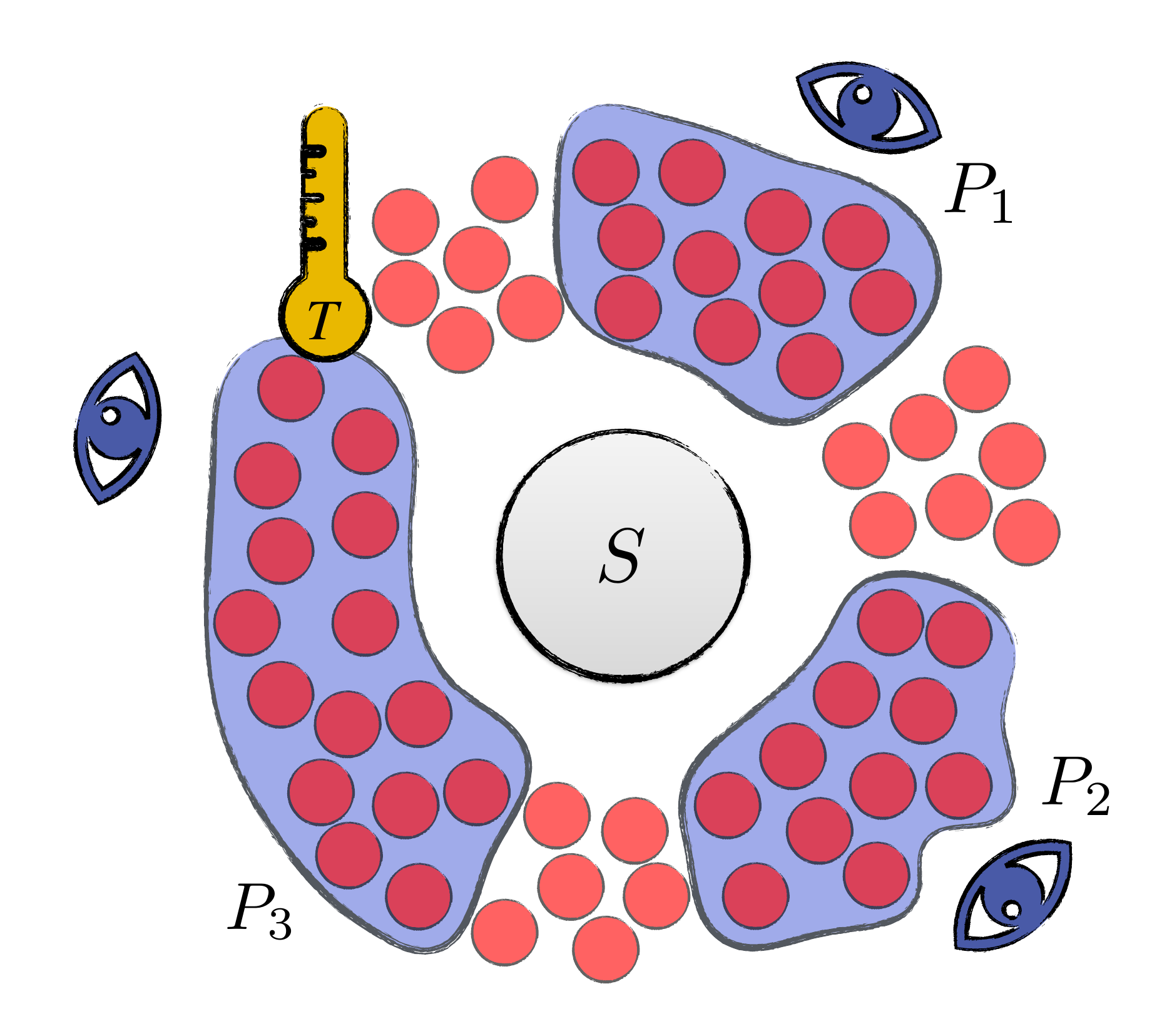}
    \caption{Multiple macroscopic observers $\mathcal{P}_j$ 
    monitor a central quantum system $S$ in order to extract information about a specific central observable. The presence of a thermal environment at temperature $T$ hinders the emergence of ideal intersubjectivity.}
    \label{fig:Fig1A}
\end{figure} 
\par 
Nonetheless, the framework developed thus far faces some limitations. First, it does not describe measurement as a process, but just as a target state, if not solely in entropic terms. Furthermore, the discussion is usually limited to projective measurement, implicitly omitting destructive measurement as a possibility. For this reason, in this article, we adopt a broader notion of objectivity known as intersubjectivity~\cite{ozawa2019intersubjectivity}. This can be understood as the minimal scenario for information broadcasting with multipartite agreement: multiple environments access local subsystems to extract information about the system's initial state. In this framework, additional properties, such as non-disturbance of the system or the environment, are not generally required. Only the broadcasting of information is essential~\cite{debarba2024broadcasting} -- i.e., the copying of the classical random variable associated with a fixed measurement. This approach thereby generalises the usual notion of ideal projective measurement and measurement schemes~\cite{busch2016quantum}, by naturally including multipartite systems as environments. 
\par 
The interplay between quantum measurement and thermodynamics has been extensively studied -- in particular, the thermodynamic cost of acquiring information through measurement~\cite{Sagawa2009,jacobs2012quantum,Faist2015a,deffner2016quantum,abdelkhalek2016fundamental,kammerlander2016coherence,linpeng2022energetic,xuereb2024resources}, and the resulting implication that limited thermodynamic resources constrain the precision with which quantum measurements can be performed~\cite{GuryanovaFriisHuber2018,debarba2019work,mohammady2023thermodynamically,panda2023nonideal,mohammady2025thermodynamic}. In this context, the measurement process has been modeled as a thermal engine, whose efficiency can be directly compared to the precision of the measurement~\cite{elouard2018efficient,latune2025thermodynamically}. Within the framework of Quantum Darwinism, the role of environmental mixedness~\cite{zwolak2009quantum}, the compatibility between equilibrium states and Spectrum Broadcast Structures (SBS)~\cite{le2021thermality}, measurement as a phase transition~\cite{allahverdyan2013understanding}, the emergence of objectivity as a classical limit of information broadcasting~\cite{debarba2024broadcasting}, and the formation of SBS through equilibration processes~\cite{schwarzhans2023,engineer2024equilibration,demelo2024finite} have been analysed.
\par 

In this work, we apply these thermodynamic considerations to characterise the emergence of intersubjectivity in a thermodynamically consistent manner (see Fig.~\ref{fig:Fig1A} for illustration). This is motivated by the observation that there is a conflict between ideal projective measurements and the third law of thermodynamics~\cite{GuryanovaFriisHuber2018}: while the measurement postulate predicts a pure post-measurement state, the third law forbids the creation of a pure -- or rank-reduced -- state with finite resources. Moreover, thermodynamics limits the ability to create the redundancies necessary to broadcast information encoded in a quantum system \cite{debarba2024broadcasting}. We investigate the constraints imposed by the third law of thermodynamics, and quantify the extent to which information can be broadcast and how consensus about observed outcomes can be achieved when available thermodynamic resources are finite. Our results thereby address the thermodynamic cost of building the type of correlations that result in intersubjectivity, adding to the broader understanding of the thermodynamic value of correlations~\cite{huber2015thermodynamic,bruschi2015thermodynamics,bakhshinezhad2019thermodynamically,piccione2020energy,piccione2021generation,krisnanda2022correlations,lipka2024fundamental,simon2025correlations,deoliveira2025heat}.
\par 
We report the following central results: first, we quantify agreement between observers and limitations on it. When the system loses its original information, we observe that interactions aimed at maximising agreement generally result in a proportional bias in the broadcasted information. Second, we discuss methods to overcome these limitations by coarse-graining environments into macrofractions. We show that the amount of achievable agreement between observers exponentially converges to unity with the size of the coarse-grained local pointers. Lastly, we compare the theoretical bound with a standard model for the emergence of objectivity: a star-spin model that results in pure dephasing. 
\par
The paper is structured as follows: in Sec. \ref{sec:intsubjOzth} we present the intersubjectivity framework; Sec. \ref{sec:3rdlawintsubj} is dedicated to the relation between intersubjectivity and the third law, proving two no-go theorems; in Sec.~\ref{sec:intsubjfires}, we quantify the deviation between imperfect and ideal intersubjectivity in terms of the thermal resources available initially, proving attainable bounds to agreement and bias; we present and discuss limits in which ideal intersubjectivity can be attained in Sec. \ref{sec:cgr} -- e.g. by coarse-graining; in Sec. \ref{sec:compstmod}, we compare our bounds to the agreement and bias in a standard model of Quantum Darwinism, finding suboptimal values for agreement and bias which can be improved by coarse-graining;  we draw our final conclusions and perspectives in Sec.~\ref{sec:conclusion}.

\section{Ideal intersubjectivity}
\label{sec:intsubjOzth}

Quantum objectivity was first formally articulated within the framework of Quantum Darwinism~\cite{zurek2003decoherence}, developed to encapsulate the essential features of the quantum-to-classical transition. More recently, the notion of objectivity has been refined into stricter formulations, such as Strong Quantum Darwinism~\cite{le2019strong} and Spectrum Broadcast Structure~\cite{horodecki2015quantum}, within an operational framework grounded in information-theoretic principles~\cite{korbicz2021roads}. However, these definitions of objectivity face two principal limitations: First, Quantum Darwinism and Strong Quantum Darwinism exclude the physically relevant case of destructive measurement, thereby imposing overly stringent conditions. Second, objectivity is characterised solely by the properties of the final output state, whereas quantum measurement should be regarded as a dynamical physical process. In this work, we address both limitations by introducing a more general, first-principles definition of objectivity, which we term \emph{ideal intersubjectivity}.
\par 
We consider the system of interest $S$ in an initial state $\rho\Sys$, and a collection of environments $\bm{P}$ which are initially uncorrelated as $\rho\PiN = \otimes_{i=1}^{\NP} \rho\Pii$. Informally, after the joint evolution of system and environments, intersubjectivity is achieved when many observers can independently access information about the initial state of the system $S$, and all observers agree on the observed outcome. In particular, the redundant encoding of the diagonal information of $\rho\Sys$ to each environmental partition $P_i$ can be considered as the broadcasting of the random variable $X$, on the system $S$, described by the probability distribution $\{p\Sys(x)\}_{x=0}^{\dS-1}$. The evolution is described by a unitary operator $U$ that yields a final state~$\trho\SPiN = U \rho\Sys\otimes\rho\PiN U^\dagger$. More formally, we require that the following three properties be satisfied for ideal intersubjectivity.\\

\textbf{Locality}: There exists a joint probability distribution among the environmental subsystems (observers) in terms of $\NP$ local measurements $\bm{\Pi}^{(i)} = \{\LocM{i}{x}\}_{x=0}^{\dS-1}$, with $i=1,...,\NP$:
    \begin{align}
        \tp\PiN(x_1,...,x_{\NP}) = \tr{\trho\PiN \bm{\Pi}_{\bm{x}}},
        \label{eq:locpr}
    \end{align}
where the measurement operators $\bm{\Pi}_{\bm{x}} = \otimes_{i=1}^{\NP} \LocM{i}{x_i}$ describe the joint measurement. The marginal probabilities
    \begin{align}
        \tp\Pii(x_i) & = \sum_{j\neq i}\sum_{x_j} \tp\PiN(x_1,...,x_{\NP}) = \nonumber \\
        & = \tr{\trho\Pii\LocM{i}{x_i}} \, ,
        \label{eq:margprobs}
    \end{align}
    correspond to local measurements by each observer.\\
    
\textbf{Probability reproducibility}: Each observer's independent measurement obeys the same probability distribution:
    \begin{align}
         \tp\Pii(x) & = \tp\Pij(x) = p\Sys(x)  \, ,
         \label{eq:probrepr}
    \end{align}
    for all $x=1,...,\dS$ and $i,j=1,...,\NP$. This property encapsulates the joint unbiasedness of information broadcasting (to local degrees of freedom).\\
    
\textbf{Agreement}: The probability of observing distinct outcomes across the different environments vanishes, i.e.,
    \begin{equation}
        \tp\PiN(x_1,...,x_{\NP}) \neq 0 \iff x_1 = ... = x_{\NP}
    \end{equation}
    Consequently, the final state of the environment $\trho\PiN$ is only supported on the agreement subspace $\mathcal{A}=\text{span}\{\otimes_{i=1}^{\NP} \Pi^{(i)}_x\}_{x=0}^{\dS-1}$. That is,
    \begin{equation}
        \sum_{x=0}^{\dS-1} \tr{\trho\PiN \otimes_{i=1}^{\NP} \Pi^{(i)}_{x}} = 1
        \label{eq:agreement}
    \end{equation}
    This condition thus ensures perfect correlations between the different environments.

Ideal intersubjectivity is defined by the simultaneous fulfilment of all three properties. Note that the second and third properties are theory-independent and can be formulated without reference to quantum mechanics. We furthermore make no assumptions regarding the dimension of the Hilbert space, and all the following results hold even in the infinite-dimensional case. The only assumption concerns the number of measurement outcomes, which must be finite, a condition justified by the finite resolution inherent to any realistic measurement procedure.

It is worth noting that the three properties so defined are not independent. Indeed, Ozawa's Theorem states that if a collection of measurements $\bm{\Pi}^{(i)}$, $i=1,...,\NP$ satisfies locality and probability reproducibility for a projective measurement on the system $S$, then it also satisfies the agreement condition~\cite{ozawa2019intersubjectivity}. 
A general proof of the theorem, which extends Ozawa's original to mixed states as required here, is provided in App.~\ref{app:ozawa_th}. The theorem implies that if information about a projective measurement on $S$ is perfectly broadcasted to multiple environments, then anyone locally accessing these systems will always agree on the observation. In other words, ideal broadcasting implies agreement and, thus, intersubjectivity. However, broadcasting redundant information cannot be achieved with constrained thermodynamic resources \cite{debarba2024broadcasting}, indicating the necessity to align with a thermodynamically consistent framework. Hence, the logical redundancy in the ideal case, as proved by Ozawa, does not directly extend to the finite resource regime, where the three properties need to be considered individually.

In this scenario, what prerequisites must be fulfilled for observers to reach \emph{agreement}, considering the possible discrepancy between the copies and the original information? Does striving for maximum agreement inherently involve reducing local bias? We address these questions in the following sections.

\section{Third law of thermodynamics and intersubjectivity}
\label{sec:3rdlawintsubj}

Ideal intersubjectivity, together with Ozawa's Theorem introduced in the previous section, provides a framework for understanding the redundant broadcasting of information from a central quantum system to multiple observers. It ensures that all observers have access to consistent information about the system and that they reach agreement on the outcome of any independent measurement run. 
\par 
On the other hand, the third law of thermodynamics implies that pure states can only be prepared at the cost of infinite resources such as time, complexity, and energy~\cite{TarantoBakhshinezhadEtAl2023}. The same holds for any state with pure subspaces -- that is, any state that is not full-rank. In ideal intersubjectivity, pure states are implicitly assumed, but, as we will argue shortly, they are necessary to satisfy the agreement condition given in Eq.~\eqref{eq:agreement}.
\par 
The agreement condition in Eq.~\eqref{eq:agreement} can only be satisfied if the final state is of sufficiently reduced rank, and such a rank-deficient state can only be unitarily reached from an initial state with that same rank. Hence, achieving ideal intersubjectivity requires starting from a severely rank-deficient initial state, a scenario that is physically unrealistic unless infinite resources are expended to cool the environments. For these reasons, thermodynamic constraints suggest that ideal intersubjective agreement cannot be exactly achieved in practice.
\par
We note here that, within thermodynamics, unitary operations describe closed systems. They represent the most natural and optimal choice among those operations that preserve rank, and thus satisfy the third law of thermodynamics.  While one might consider more general evolutions, such as completely positive trace-preserving (CPTP) maps that reduce rank, such operations are inconsistent with thermodynamic principles, as they explicitly violate the third law, any unitary dilation in turn requiring rank-deficient input states. Consequently, we restrict our attention to evolutions that do not decrease rank, with unitary dynamics being the most efficient within this class. Nevertheless, our approach is also consistent with more general non-unitary thermal operations~\cite{Janzing2000,Brandao2013}, as auxiliary systems can be coupled to each pointer and subsequently discarded during the readout of the statistics.
\par
The impossibility of achieving agreement under full-rank state preparation, as dictated by the third law, has deeper implications for the framework of intersubjectivity. Ozawa's Theorem asserts that locality and probability reproducibility together imply agreement. Thus, the logical negation of this statement indicates that if agreement cannot be achieved, then either locality or probability reproducibility must be abandoned. Since locality is essential for intersubjectivity, allowing multiple observers to access information locally from their respective environments, we are forced to abandon or at least weaken the requirement of probability reproducibility, and, by extension, ideal classical broadcasting. We summarise these results in the following observation
\par 

\begin{obs}[Impossibility of thermodynamically-consistent ideal intersubjectivity]
    \label{lem:no-go-ideal-int}
    Agreement (Eq. \eqref{eq:agreement}) and probability reproducibility (Eq. \eqref{eq:probrepr}) cannot be satisfied simultaneously, and ideal intersubjectivity cannot be attained without the use of diverging thermodynamic resources.
\end{obs}

\par
This result raises an important question: given a finite amount of resources, how closely can we approximate ideal intersubjectivity? In the next section, we address this question in depth, describing the fundamental limits of finite resources to agreement and probability reproducibility.

\section{Finite resource-consistent intersubjectivity}
\label{sec:intsubjfires}

As ideal intersubjectivity is in conflict with the third law of thermodynamics, we now turn to investigating how closely it can be approximated. Recalling the definitions of local measurement probabilities in Eqs.~\eqref{eq:locpr}~and~\eqref{eq:margprobs}, we start by introducing a more permissive version of intersubjectivity, which we refer to as \emph{biased joint information broadcasting}:
\begin{definition}[Biased joint information broadcasting (BJIB)]
    A process leads to biased joint information broadcasting if many observers can access the same information locally, reproducing a noisy version of information about the initial state of the system $S$:
    \begin{equation}
        \tp\Pii(x) = \tp\Pij(x) \neq p\Sys(x)\, ,
        \label{eq:bjb}
    \end{equation}
for all $x=0,...,\dS-1$ and for all $i,j=1,...,\NP$. 
\label{def:bjib}
\end{definition}
This definition agrees with semi-classical information broadcasting \cite{debarba2024broadcasting}, where the information accessed locally is imperfect with respect to the one originally intended to be copied. Indeed, Eq. \eqref{eq:bjb} can be understood as a more permissive and biased version of probability reproducibility in Eq.~\eqref{eq:probrepr}.

As stated in Observation~\ref{lem:no-go-ideal-int}, the ability to reproduce redundant information is naturally constrained by the thermodynamic resources in the environments. Here, a finite-resource quantum thermodynamic process is characterised by the unitary interaction of the system state $\rho\Sys$ (on Hilbert space $\mathcal{H}\Sys$) with a thermal environment prepared as 
\begin{equation}\label{gibbs_state}
    \tau_\beta = \frac{e^{-\beta H}}{Z_\beta}\, ,
\end{equation}
with inverse temperature $\beta$, Hamiltonian $H=\sum_i E_i\vert E_i\rangle \langle E_i \vert$, collective environmental energy levels $E_i$, and partition function $Z_\beta = \tr{e^{-\beta H}}$. For non-interacting environmental subsystems, the set $\{\ket{E_i}\}$ includes all possible combinations of local environmental energy eigenstates. At non-zero temperature, the state $\tau_\beta$ is full-rank.
\par 
As the information about the system needs to be encoded in subspaces of the environment, we must decompose each local Hilbert space into $\dS$ subspaces $\text{span}\{\vert E_i \rangle \langle E_i\vert\}_{i \in D_x}$, where $\dS$ is the number of outcomes, $D_x$ is a possibly infinite collection of indices that specify the orthognal subspace corresponding to the $x$th outcome.  In this way, we can rewrite our initial environment state as
\begin{equation}
    \tau_\beta = \sum_{x=0}^{\dS-1} A_x\, ,
\end{equation}
with each orthogonal component $A_x$ defined as
\begin{align}
    A_x  =& \sum_{i \in D_x} \frac{e^{-\beta E_i}}{Z_\beta} \vert E_i \rangle \langle E_i \vert \,.
    \label{eq:ax_mat_def}
\end{align}
We denote the trace of these orthogonal components as $\ax{x} = \tr{A_x}$, and we also define $\bm{a} = \{\ax{0},\ax{1},\ldots,\ax{d\Sys -1}\}$. This decomposition is, in principle, arbitrary, but in the following, we will show how to select and identify the one that allows for optimal intersubjectivity with finite resources. We stress that we have not assumed that the environment has a finite-dimensional Hilbert space, only that the number of outcomes is finite. 

We proceed to quantify how close BJIB can get to ideal intersubjectivity. To do so, we define two quantities that capture the deviation from ideal intersubjectivity of a unitary process $U$. First, for the final state $\trho \SPiN = U \rho\Sys\otimes\rho\PiN U^{\dagger}$, with the composed pointer prepared as $\rho\PiN = \tau_{\beta}^{\otimes \NP}$, the \textit{agreement} is quantified by
\begin{align}
    \Agr{U} & = \sum_{x=0}^{\dS-1} \tr{\trho\SPiN \id \bigotimes_{i=1}^{\NP} \Pi^{(i)}_{x}} \label{eq:agr_bjb}\, .
\end{align}
This quantity measures the support of the final state on the agreement subspace $\mathcal{A}$. The value of $\Agr{U}$ reflects the probability on average of the observers to agree on their measurement output.  Conversely, \textit{disagreement} is naturally given by $\Dis{U} = 1 - \Agr{U}\,$.
\par 
We now have all the ingredients for introducing our main result: 
\begin{theorem}[Maximum agreement]\label{th:fres-intsub}
For biased joint information broadcasting, as defined in \ref{def:bjib}, the maximal agreement is
     \begin{equation}
         \max_U \Agr{U} := \fagr{\NP} = \sum_{x=0}^{\dS-1}\ax{x}^{\NP}\,.
         \label{eq:MaxAgrB}
     \end{equation}
     and there exists an evolution $\UBJB{opt}$ attains this maximum.
\end{theorem}

Any other unitary $U$ that satisfies BJIB will have a smaller agreement, i.e., $\Agr{\UBJB{}} \leq \fagr{\NP}$. The maximally attainable value is in turn definitionally upper-bounded by unity: $\fagr{\NP}\leq 1$.

The formal proof of Thm.~\ref{th:fres-intsub} is provided in App.~\ref{app:proof-fres-intsub}, with an explicit construction of the optimal unitary which makes Eq.~\eqref{eq:MaxAgrB} achievable. We sketch its essence here: this relies on the decomposition of the initial state of the environment between the agreement subspace $\mathcal{H}_{\mathcal{A}}$ and its orthogonal part, the disagreement subspace $\mathcal{H}_{\mathcal{D}}$. In this way, the initial state becomes
\begin{equation}
    \rho\PiN = \rho^{\text{agr}}\PiN + \rho^{\text{dis}}\PiN \,.
\end{equation}
Now, to have a maximal agreement, we need to focus on how to realise an unbiased information broadcast of the probability $\{p\Sys(x)\}_x$ with respect to the agreement subspace. This constraint the unitary evolution on $\mathcal{H}\Sys\otimes \mathcal{H}_{\mathcal{A}}$. To fix the unitary also on the disagreement subspace, we enforce the biased joint information broadcast condition in Eq. \eqref{eq:bjb}. This additionally forces the unitary evolution to be a specific form, eventually fixing the overall unitary evolution on the total Hilbert space.

Theorem~\ref{th:fres-intsub} establishes an achievable upper bound on the maximum agreement given a fixed amount of resources, as captured by the quantity $\fagr{\NP}$. The maximal agreement depends only on this function, which is independent of the initial state of the system $\rho\Sys$ while being only a function of the initial state of the environments $\rho\PiN$ and on the number of environments $\NP$ that must agree. One can note that $\sum_{x=0}^{\dS-1}\ax{x} \equiv \tr{\tau_\beta}=1$, therefore it will only be equal to one, satisfying the ideal intersubjectivity if each local pointer's initial state has a single support subspace with dimension smaller than $d\Poi/\dS$, in other words $\text{rank}(\rho\Poi)< d\Poi/\dS$. This implies that each individual environment does not need to cool down its local pointers to a rank-$1$ state to obtain perfect intersubjective broadcasting. However, it must be cold enough to have a specific support subspace that fits all the system information. Furthermore, one can define the minimum observable disagreement as $\fdis{\NP} = 1 - \fagr{\NP}$.

\par 
A comment on the dynamics that yield the maximum agreement is also necessary. Theorem~\ref{th:fres-intsub} above provides a class of unitaries that realise the maximum agreement. An underlying Hamiltonian description can be obtained by taking the logarithm of the unitary operator. Rather than further relating such a procedure to basic, physical properties, we here give a simple circuit-inspired implementation of the required unitary itself: this can be achieved by applying a generalised SWAP gate between the system and one environment and then applying a sequence of generalised CNOT gate on the environments.
\par
According to Observation~\ref{lem:no-go-ideal-int}, the information locally encoded in the environments will be biased, as the information cannot be completely copied to each pointer. The local bias can be quantified by the statistical distance between the probability distribution obtained by a local observer $\tp\Pij$ and the probability distribution $p\Sys$ on the initial state of $S$ that we originally aimed to broadcast, 
\begin{align}
    \Biasj{U}{j} & = D_{\mathrm{T}}(\tp\Pij,p\Sys)
    \label{eq:bia_bjb}
\end{align}
where $D_{\mathrm{T}}(p,q)=\sum_i|p_i - q_i|/2$ refers to the $l_1$-distance between $\{p(x)\}$ and $\{q(x)\}$. We note that for systems satisfying BJIB, the bias will not depend on $j$, so we drop it from our notation in the remainder of the paper and simply use $\Bias{U}$. 
\par 
In Theorem \ref{th:fres-intsub}, we have only discussed the maximum agreement. We now address the bias of such optimal unitary that we have found, and summarise the result in the following theorem:
\begin{theorem}[Bias for optimal BJIB] \label{th:fres-biasBJIB}
     For the optimal unitary $\UBJB{opt}$, the local probabilities can be written as a convex combination of the original probability and a noisy contribution
     \begin{equation}
         \tp\Pii(x) = \fagr{\NP} p\Sys(x) + (1-\fagr{\NP}) \mst{x}\, ,
     \end{equation}
    where
     \begin{equation}
         \mst{x} = \frac{\ax{x}(1-\ax{x}^{\NP-1})}{1-\fagr{\NP}}\, .
     \end{equation}
     As a result, the bias is
     \begin{equation}
         \Bias{\UBJB{\text{opt}}} = \fdis{\NP} D_{\mathrm{T}}(p\Sys(x),\mst{x})
         \label{eq:BiaOptU} := \beta_{\bm{a}}^{(\NP)}\,.
     \end{equation}
\end{theorem}
The detailed proof is provided in App.~\ref{app:proof-fres-biasBJIB}. Here, one can observe that the bias of the optimal unitary depends on the probability distribution to be copied $\{p\Sys(x)\}_{x}$. On the other hand, it is proportional to the minimum disagreement, defined as $\fdis{\NP} = 1 - \fagr{\NP}$. Consequently, as this quantity approaches zero, both perfect agreement and vanishing bias are attained, thereby realising ideal intersubjectivity. 
\par 
Furthermore, we note that minimum disagreement $\fdis{\NP}$ can also be written as
\begin{equation}
    \fdis{\NP} = (\NP-1) S_{\NP} (\bm{a})\, ,
\end{equation}
where
\begin{equation}
    S_q(\bm{p}) = \frac{1-\sum_i p_i^q}{q-1} \, ,
\end{equation}
is the Tsallis entropy \cite{tsallis1988possible,tsallis2011nonadditive}, which has applications in systems where energy and entropy are non-extensive \cite{krisut2024deriving}. We thus conclude that the disagreement is proportional to $\NP$ and the Tsallis entropy of $\bm{a} = \{\ax{0},\ax{1},\ldots,\ax{d\Sys -1}\}$. Note that for a pure state of the environment (i.e., at zero temperature), the agreement converges to one, and the minimal bias diminishes to zero, as expected.

As the number of environments $\NP$ increases, the upper bound $\fagr{\NP}$ correspondingly decreases, reflecting the increased difficulty of establishing perfect correlations among a larger number of systems under fixed thermodynamic resource constraints. We also note that as $\fagr{\NP}$ gets closer to $1$, bias goes to zero, achieving ideal intersubjectivity. Conversely, for all non-optimal unitaries in Eq. \eqref{eq:bjb}, the agreement is smaller. 

\par 
As previously noted, enhancing agreement and reducing bias requires increasing the value of $\fagr{\NP}$ as close to unity as possible. This can be achieved by appropriately selecting the dimensions and components of the subspaces $D_x$, as discussed in the text above. The optimal configuration can be constructed by the following procedure: first, assign the eigenvectors $\vert E_i\rangle\langle E_i\vert$ associated with the smallest eigenvalues $E_i$ to the largest subspace $D_x$, where “largest” refers to the subspace with the greatest dimension $d_x$. Next, assign the eigenvectors corresponding to the second-smallest eigenvalues to the second-largest subspace, and so on. By sequentially filling all subspaces in this manner, we maximise $a_0$, then $a_1$ (given the maximal $a_0$), and so on, ultimately yielding the highest possible value for $\fagr{\NP}$. Notably, the grouping ${D_x}$ is independent of $\NP$. This optimal arrangement can be realised when there is sufficient freedom to design the local measurement $\{\LocM{i}{x}\}_{x=0}^{d\Sys-1}$. However, in the presence of physical constraints, one may be limited to performing a fixed measurement $\{\Pi_x\}_x$. As a result, such restrictions also limit the ability to optimise over the choice of ${D_x}$, thereby imposing a corresponding constraint on the attainable value of $\fagr{\NP}$.
\par 
Following this initial optimisation step, the most natural strategy to further increase $\fagr{\NP}$ is to increase the purity of the initial environmental states $\rho\Pii$. However, this approach may be unfeasible if the environments are insufficiently isolated or if the energy required is prohibitively high. For these reasons, in the next section, we introduce an alternative method to improve $\fagr{\NP}$ without increasing the amount of initial resources.

\begin{figure}
    \centering
\includegraphics[width=0.95\linewidth]{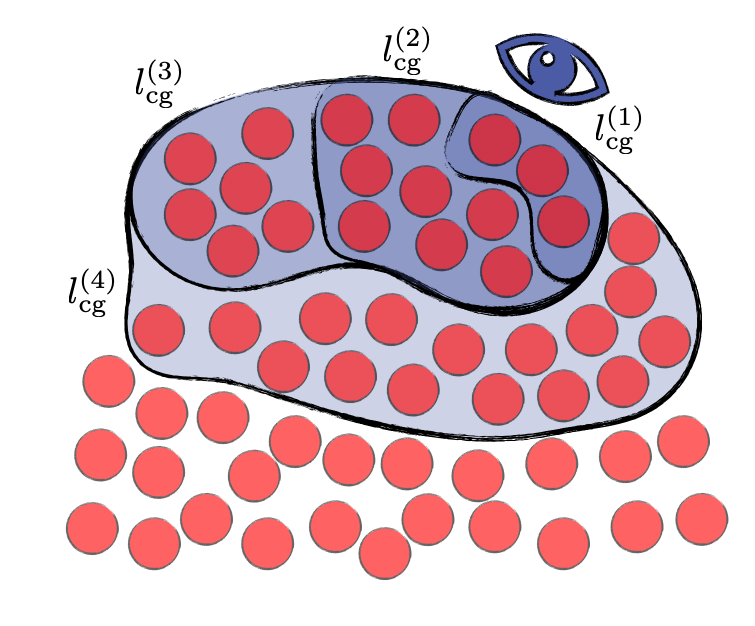}
    \caption{Representation of coarse-graining: The environments are grouped into larger and larger macrofractions of size $\lcgu$, from which information is extracted.} 
    \label{fig:Fig1B}
\end{figure}

\section{Coarse-graining into macrofractions}\label{sec:cgr} 
\par
The coarse-graining of the environments has been used to approach ideal objectivity, for instance, in Refs.~\cite{zurek2003decoherence,korbicz2021roads}. Here, coarse-graining refers to the grouping of subsystems into a larger composite system, commonly referred to as macrofractions~\cite{blume2006quantum,mironowicz2017monitoring}. We show that with this method, we can approximate ideal intersubjectivity even with a finite amount of thermodynamic resources. This is a feasible method for increasing $\fagr{\NP}$ once the initial states and purities of the environments are fixed, thereby establishing coarse-graining as a necessary condition for approaching ideal intersubjectivity. Our result generalises previous model-dependent findings~\cite{mironowicz2018system,PhysRevA.109.032221,lee2024encoding,schwarzhans2023}, highlighting the essential role of coarse-graining in attaining ideal intersubjectivity within a model-independent framework.
\par
We identify two different methods of coarse-graining. In the first case, we coarse-grain the system by simply incorporating subsystems together and keeping the total number of subsystems $\NP$ constant. This method simply increases the size of the environments and, as we will show, makes $\fagr{\NP}$ larger. This approach is particularly well-suited for infinite-dimensional systems, such as bosonic systems, where analysis is restricted to a finite subset of modes. We adopt a similar strategy in the second case but maintain a fixed total Hilbert space dimension. In doing so, the grouping of subsystems into macrofractions effectively reduces the total number of environments, i.e., $\NP \to \NPcg$. See Fig.~\ref{fig:Fig1B} for a pictorial representation. 
\par 
Whether we use the first or second coarse-graining method, coarse-graining corresponds to a mapping of the original environments $\{P_i\}_{i=1}^{\NP}$ into a new set of coarse-grained environments $\{P_i^{\lcg}\}_{i}$. The total number of environments after coarse-graining depends on the specific method used. However, regardless of the method, in our notation, each coarse-grained environment consists of $\lcg$ original environments grouped together. Following this redefinition, the initial environmental state is replaced by $\tau_\beta^{\otimes \lcg}$, and consequently, the decomposition introduced in Eq.~\eqref{eq:ax_mat_def} must also be updated accordingly. This results in the coarse-grained versions $A_{x}^{\lcg}$ and $\axl{x}{\lcg}$. Formally, when we coarse-grain, the initial coarse-grained environment state is
\begin{equation}
    \rho\Pilcg = \bigotimes_{j \in D^{\lcg}_i} \rho\Pij = \sum_{x_1,...,x_{\lcg}} A_{x_1} \otimes ... \otimes A_{x_{\lcg}}\, ,
    \label{eq:cg-state}
\end{equation}
where $D^{\lcg}_i$ is a collection of indices corresponding to systems constituting the $i$th macrofraction. The dimension of the macrofraction is $\lcg = \vert D^{\lcg}_i \vert$, which we assume to be independent of $i$, since we consider all new $\Pilcgntext$ to have the same dimension.
\par 
We now want to decompose this state as
\begin{equation}
    \rho_{\Pilcg} = \sum_{x} A^{\lcg}_{x} 
\end{equation} 
in such a way that $\axl{x}{\lcg} = \tr{A^{\lcg}_x}$ increases the coarse-grained optimal agreement $\fagrl{\NP}{\lcg}$, as we increase $\lcg$. Regardless of the coarse-graining method applied, the problem simply consists in understanding how to construct $A^{\lcg}_x$ (or $\axl{x}{\lcg}$), from combinations of $A_{x_1}\otimes..\otimes A_{x_d}$ (or $\ax{x_1}...\ax{x_d}$). After this reconstruction, the original bound $\fagr{\NP}$ gets remapped to $\fagrl{\NP}{\lcg} := \sum_x \axl{x}{\lcg}^{\NP}$, along with $\fdisl{\NP}{\lcg}$ and $\beta^{(\NP)}_{\bm{a},\lcg}$.
\par  
In general, there is no unique way to employ such mapping from original environments to coarse-grained ones. However, for $\dS=2$, the following scaling holds
\begin{theorem}[Approaching ideal intersubjectivity with coarse-graining]\label{th:idealintcg}
For $\dS=2$ outcomes, there exists a coarse-graining for which the function $\fagrl{\NP}{\lcg}$ scales as 
\begin{equation}
    \fagrl{\NP}{\lcg} \simeq 1 - \NP\frac{e^{-D(\bm{a})(\lcg-1)}}{\sqrt{\lcg}} F(\bm{a}) + O(\axl{\lcg}{1}^2)\, ,
\end{equation}
for $\lcg$ large. Here, $D(\bm{a})$ and $F(\bm{a})$ are functions that depend solely on $\bm{a} = \{\ax{0},\ax{1},\ldots,\ax{d\Sys -1}\}$ and not on the size of the coarse-graining $\lcg$.
\end{theorem}
This theorem is demonstrated in App.~\ref{app:coarse-grain-proof}, where we also provide a precise definition of the coarse-grained projectors $\{A^{\lcg}_x\}_x$. Notably, $\fagrl{\NP}{\lcg}$ converges exponentially to $1$ as a function of $\lcg$, indicating that ideal intersubjectivity can, in general, be recovered even under finite resources, provided that subsystems are grouped into sufficiently large macrofractions.
\par 
For $\dS > 2$, deriving an analytical formula becomes particularly challenging. Therefore, to determine whether $\fagrl{\NP}{\lcg}$ approaches $1$ exponentially fast as $\lcg$ increases, we numerically explore the scaling of $\fagrl{\NP}{\lcg}$. More precisely, we investigate the behaviour of $\axl{\lcg}{0}$, that is, how the largest probability scales. In fact, to guarantee that $\fagrl{\NP}{\lcg}$ increases with $\lcg$ and approaches $1$, it is sufficient for $\axl{\lcg}{0}$ to also increase with $\lcg$ and to converge toward $1$.  
\par
\begin{figure}[t]
    \centering
    \begin{minipage}{0.49\textwidth}
        \centering
        \includegraphics[width=\textwidth]{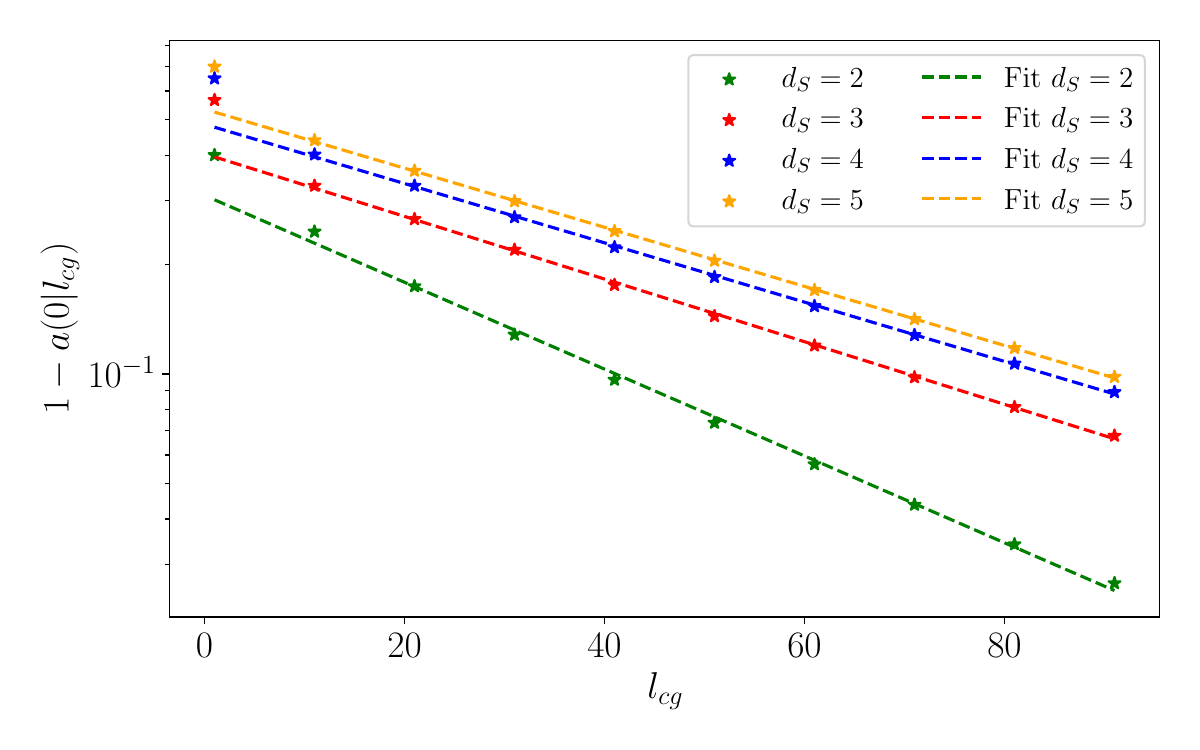}
    \end{minipage}
    \hspace{0.02\textwidth} 
    \begin{minipage}{0.4\textwidth}
        \centering
        \begin{tabular}{cccc} 
        \toprule
        $\dS$ & $R^2$ & $c_0$ & $c_1$ \\
        \midrule
        2  &  0.99726  & 0.30992 & -0.02747 \\
        3  &  0.99943  & 0.40351& -0.01982 \\
        4  &  0.99966  & 0.48630 & -0.01875 \\
        5  &  0.99988  & 0.53518 & -0.01869\\
        \bottomrule
        \end{tabular}
    \end{minipage}
    \caption{In this figure we plot the exponential decay of $1-\axl{0}{\lcg}$ to extend the result of Theorem \ref{th:idealintcg}. Top panel: Logarithmic plot of $1-\axl{0}{\lcg}$ as a function of~$\lcg$. Stars correspond to values evaluated with the definition in App. \ref{app:coarse-grain-proof-dslarger}; Dashed lines correspond to fit with $c_0 e^{c_1 \lcg}$. Table: Results of the fit. We assess the efficiency of the fit with the coefficient of determination $R^2$, see App.~\ref{app:coarse-grain-proof-dslarger} for details. We considered initial probability $\bm{a}_{\dS} =  \{\ax{0} = 1/\dS + 0.1, \ax{1}=..=\ax{\dS-1} = 1/\dS-0.1/(\dS-1) \}$, which is close to a maximally mixed initial configuration. The fit is performed for $\lcg > \lcg^{\text{min}}$ to demonstrate the asymptotic behaviour. In practice, we excluded the first point $\lcg=1$.}
    \label{fig:ax0lcg}
\end{figure}

 
Inspired by the case $\dS=2$, we devise a similar strategy to group products of $\ax{x_i}$. We refer the reader to App.~\ref{app:coarse-grain-proof-dslarger} for details. In Fig.~\ref{fig:ax0lcg}, we show $1-\axl{\lcg}{0}$ as a function of $\lcg$. A fit of the form $c_0 e^{c_1 \lcg}$ is also provided, along with the fit parameters listed in Table \ref{fig:ax0lcg}. We observe that the model provides a good fit, and we argue that $\fagrl{\NP}{\lcg}$ approaches 1 exponentially fast with respect to $\lcg$, even for $\dS > 2$. For completeness, we also include the case $\dS = 2$ in the plot. Additionally, we note that the rate of the exponential decay remains nearly constant across the cases $\dS > 2$, indicating that the functional dependence of $c_0$ and $c_1$ does not vary significantly.
\par 
We conclude this section with a remark regarding our result. In this section, we showed that coarse-graining, if properly performed as described in App. \ref{app:coarse-grain-proof}, can yield ideal intersubjectivity even with finite resources in a model-independent fashion. No other assumptions were made in that regard, apart from unitary evolution. We show in the next section that intersubjectivity can be achieved even without ideal unitary evolution. We conclude that coarse-graining proves to be a sufficient condition to attain ideal intersubjectivity. On the other hand, there are no other strategies that we can devise other than grouping environments together to increase $\fagr{\NP}$. It thus becomes a necessary condition as well, without requiring full cooling of all pointers and the consequent use of infinite thermodynamic resources.

\section{Comparison with the standard model of objectivity}
\label{sec:compstmod}

\begin{figure*}
    \centering
    \includegraphics[width=1\linewidth]{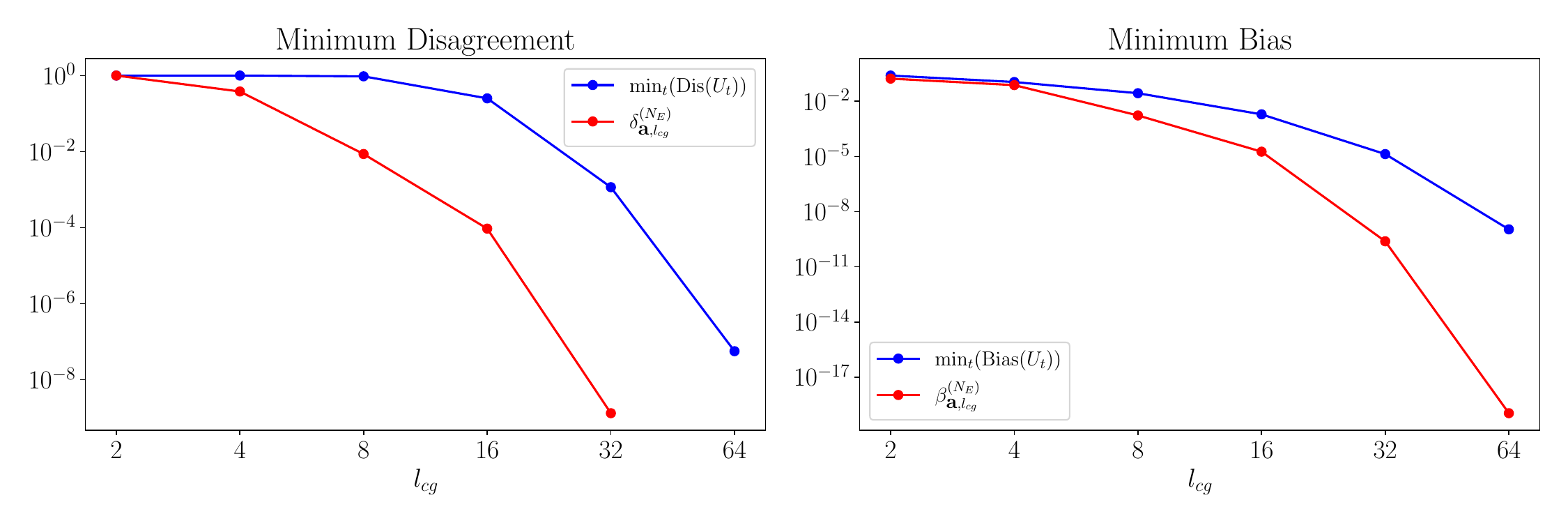}
    \caption{We report minimum disagreement $\textup{Dis}(U_t)$ and bias $\textup{Bias}(U_t)$ on the left and right canvas, respectively, for a pure dephasing dynamics generated by a star shape spin interaction, see Eq. \eqref{eq:Hspin}. The quantity are evaluated for initial state $\vert \psi_{0}\rangle = \sqrt{p_{0}} \vert 0 \rangle + \sqrt{1-p_{0}} \vert 1 \rangle$ with $p_{0}=0.2$, while the environments are in a thermal state with inverse temperature $\beta=1$ and the total number of environmental qubits is $\NP = 1024$. We evaluate the minimum of the two considered quantities over the time interval given as $t \in[0,6]$, and compare them with the corresponding coarse-grained version of the bound, namely $\fdisl{\NP}{\lcg}$ and $\beta_{\bm{a},\lcg}^{(\NP)}$. We observe a similar behaviour in both the bound and pure dephasing models, with the only distinction being that the decrease is slower in the pure dephasing case. Nevertheless, it can still be characterized as an exponential decay. Note that the axis uses an exponential scaling in the $\lcg$. For $\lcg = 64$, the bound for $\fdisl{\NP}{\lcg}$ was too small to be represented in the plot.}
    \label{fig:dis-bias-pure-deph}
\end{figure*}
Most of the results in Quantum Darwinism and objectivity are Hamiltonian-dependent. In this paper, however, we have established some general results using an approach that does not rely on a specific model and holds for arbitrary unitaries. In this section, we aim to connect our theoretical findings to a standard model of objectivity in order to show how agreement and bias in a specific model relate to our bounds.
\par
We consider a central qubit system $S$ coupled to $\NP$ qubits. The resulting so-called pure-dephasing Hamiltonian can be written as
\begin{equation}
    \HPD = \frac{1}{2} \sum_{i=1}^{\NP} \sigma_z^{(\Sysu)} \otimes \sigma_z^{(i)} \, ,
    \label{eq:Hspin}
\end{equation}
with $\sigma_z^{(\Sysu)}$ the Pauli operator acting on the system, while $\sigma_z^{(i)}$ acting on $P_i$ only. This model had been extensively studied in the literature \cite{zwolak2009quantum,zwolak2010redundant,mironowicz2017monitoring}, proving to result in quantum Darwinism and SBS structure in the coarse-grained limit.
\par 
The initial state of the system is given by $\vert \psi \rangle = \sqrt{p_0} \vert 0\rangle + \sqrt{1-p_0} \vert 1\rangle$, while each environment is initialised in a thermal state diagonal in the $\sigma_x^{(\Piiu)}$ basis. The joint evolution under the unitary $U_{\star}(t)$, generated by the Hamiltonian in Eq.~\eqref{eq:Hspin}, leads to the final state $\rho\SPiN^t$. To extract the information that has been broadcast to the environments, we employ a local measurement $\{\LocM{i}{x}\}_{x=0,1}$ on each pointer, optimised to distinguish between the evolved branches of the system state. An analogous procedure is used in the coarse-grained scenario, where the measurement is now defined on the macrofractions, i.e., $\{\LocM{i,\lcg}{x}\}_{x=0,1}$.
\par
To simulate the evolution with a very large number of environmental spins $\NP$ and large macrofractions, we employ a method developed in Ref.~\cite{zwolak2010redundant}. This approach takes advantage of the fact that dynamics generated by Eq.~\eqref{eq:Hspin} can be efficiently described in terms of spin states, significantly reducing the numerical memory required for the simulation. Further details are provided in App.~\ref{app:ev-puredephH}, while here, we only provide a sketch of the method used.
\par 
Given the structure of the Hamiltonian, the unitary evolution can be written in terms of a conditional unitary
\begin{equation}
	U_{\star}(t) = \vert0\rangle\langle 0 \vert \otimes \left[V(t)^{\otimes \NP}\right] + \vert1\rangle\langle 1 \vert \otimes \left[V(-t)^{\otimes \NP}\right] \, 
\end{equation}
which results in an environmental state at time $t$ that reads as
\begin{equation}
    \rho\PiN(t) = p_0 \tilde\rho_0(t)^{\otimes \NP} + p_1 \tilde\rho_1(t)^{\otimes \NP}\, ,
\end{equation}
where $\tilde\rho_0(t) = V(t) \rho_{\Poi} V(t)^\dagger$ and $\tilde\rho_1(t) = V(-t) \rho_{\Poi} V(-t)^\dagger$ are the local environments for the respective branch. The decomposition of the state into spin states allows us to represent the system in a block diagonal form. Further, for a macrofraction of size $\lcg$ we have that
\begin{equation}
	\rho\Pilcg(t) = \bigoplus_{j=0}^{\lcg/2} \left[p_0 N^{(0)}_j(t) + p_1 N^{(1)}_{j}(t)\right]^{\oplus B_j}\, ,
\end{equation}
see App.~\ref{app:ev-puredephH} for details and definitions of $N_{j}^{(x)}$ and $B_{j}$. Here $N_{j}^{x}$, $x=0,1$ are matrices whose dimension is up to $(2 \lcg +1) \times (2\lcg +1)$, which means that we need to evaluate matrices with dimensions that scale linearly with $\lcg$ rather than exponentially. 
\par
Using this structure, we can efficiently simulate the evolution of the total environment and evaluate the optimal measurement for discriminating between the two branches based on optimal discrimination theory \cite{bae2015quantum, bergou2010discrimination}. Notably, the optimal discrimination measurement can also be expressed in a block-diagonal form, further enhancing computational efficiency.
\par 
We simulate a pure-dephasing model with up to $1024$ environmental spins and evaluate agreement and bias, respectively, defined in Eq.~\eqref{eq:agr_bjb} and Eq.~\eqref{eq:bia_bjb}, and compare it with respect to our bounds. More specifically, we compute the coarse-grained version of agreement and bias as a function of $\lcg$ and evaluate their minimum on a typical timescale of the Hamiltonian evolution. We compare these values to the coarse-grained version of our bounds, namely $\fagrl{\lcg}{\NP}$ and $\beta_{\bm{a},\lcg}^{(\NP)}$. We report Disagreement and Bias in Fig.~\ref{fig:dis-bias-pure-deph}: we observe that, as the coarse-graining size $\lcg$ increases, not only does $\fdisl{\NP}{\lcg}$ approach zero -- consistent with the results of the previous section -- but both the minimum disagreement for the pure-dephasing model and the bias also exhibit an exponential decay toward zero, albeit at a slower rate.  
\par 
We also observe that there is always a gap between our bound and the agreement and bias evaluated for the pure dephasing Hamiltonian. This discrepancy is mainly due to the fact that the pure dephasing evolution described by the Hamiltonian in Eq.~\eqref{eq:Hspin} corresponds to a local interaction. In contrast, the optimal unitary evolution might not be local and could require additional fine-tuning. However, the values ultimately approach $0$ as the size of the coarse-graining increases. This indicates that locality plays a role in determining the rate at which we approach ideal intersubjectivity, but not whether ideality is ultimately attained by coarse-graining. This further suggests that coarse-graining methods are crucial for universal objectivity, regardless of the model used.
\par 
As a final comment on the central spin model, we emphasise that the coarse-graining method also influences the agreement time, i.e., the minimum time required to reach an $\epsilon$-close approximation to the ideal values of bias and agreement. In App.~\ref{app:cg-time}, we plot the agreement and bias as functions of time for various values of $\lcg$. While the maximum (minimum) agreement (bias) stabilises beyond a certain coarse-graining size, increasing $\lcg$ still has a significant effect on the time required to reach these extremal values (see Fig.~\ref{fig:pure-deph-agr-bias}). Specifically, larger coarse-graining sizes lead to a reduction in the minimum time needed to achieve the desired level of precision. Intuitively, this is due to the increasing Hilbert space dimension allocated the output subspaces, thereby making different branches easier to distinguish between one another and more quickly so as the joint system evolves. This suggests that coarse-graining impacts not only the asymptotic values of agreement and bias but also the dynamics of convergence, potentially accelerating the measurement process -- a direction we leave open for future investigation.

\section{Discussions and Conclusions}
\label{sec:conclusion}
The interconnection between measurement, thermodynamics, and the emergence of classicality has been of central research interest in quantum thermodynamics and the foundations of quantum theory. Particularly the clash between textbook projective measurement and the third law of thermodynamics has received attention in recent years. We address this question by appealing to quantum intersubjectivity -- a more permissive form of objectivity that does not require the system to be preserved when observers gather information about it. Here, ideal intersubjectivity is defined by the requirements of locality, probability, reproducibility, and agreement (between different observers). Ozawa's Theorem~\cite{ozawa2019intersubjectivity} revisited in Sec.~\ref{sec:intsubjOzth} and in App.~\ref{app:ozawa_th} established agreement as a necessary result of the other two properties.

We first showed that ideal intersubjectivity cannot be unitarily produced between observers that are initially in local thermal equilibrium (Observation 1). This first observation can be directly intuited from the fact that agreement implies a rank-reduced final state compared to the generally full-rank initial state. It naturally leads to the questions of how close we can get to ideal intersubjectivity, if it can be approximated, and at what cost.

Following this observation, our first main result (Theorem~\ref{th:fres-intsub}) gives a saturable upper bound for the maximal amount of agreement that can be achieved with an initially thermal environment. While the optimal unitary operation that does so is not unique, an explicit construction is provided in App.~\ref{app:proof-fres-intsub}. The second main result was then to show the unavoidable trade-off incurred when optimising for agreement: a bias is introduced vis-à-vis the original statistics encoded in the system. Thm.~\ref{th:fres-biasBJIB} quantifies this bias for the case of a maximal-agreement unitary. As the last of our three main results, we showed that coarse-graining (i.e., combining parts of the environment into larger macro-environments) lets us recover perfect agreement arbitrarily closely. This is stated precisely in Thm~\ref{th:idealintcg}. Therefore, ideal intersubjectivity is recovered not only in the limiting case of cooling the environment to absolute zero but also at finite temperatures, provided the environment undergoes appropriate coarse-graining. This finding reconciles Ozawa's conclusions with the conventional macroscopic limit, here formalised as a coarse-graining procedure essential for a consistent quantum-to-classical transition. It parallels a similar recent use of coarse-graining for obtaining objectivity under equilibration of a closed quantum system~\cite{schwarzhans2023}. We finished by illustrating our results in the relevant limits of the paradigmatic central-spin system in Sec.~\ref{sec:compstmod}.

Extending this work, it would be interesting to investigate how the emergence of intersubjectivity is further constrained by considering a smaller set of processes, such as the ones due to microscopic energy-conservation and thermal operations~\cite{Janzing2000,Brandao2013,ng2019resource,lostaglio2019introductory}.

Finally, we observed that the speed of measurement, defined as the time scale at which we approach ideal intersubjectivity, strongly depends on the level of coarse-graining applied to the environments. Quantifying this dependence constitutes a tantalising avenue for future research.

\vspace{0.5cm}
\begin{acknowledgments}
The authors are grateful to Jarosław K. Korbicz and Nicolai Friis, for fruitful discussions. 

This publication was made possible through the support of Grant No. 62423 from the John Templeton Foundation. The opinions expressed in this publication are those of the authors and do not necessarily reflect the views of the John Templeton Foundation. 

A.C. acknowledges support from the “Italian National Quantum Science and Technology Institute (NQSTI)” (PE0000023) - SPOKE 2 through project ASpEQC.
T.D. acknowledges support from {\"O}AW-JESH-Programme; from the European Research Council (Consolidator grant ’Cocoquest’ 101043705) and the Brazilian agencies CNPq (Grant No. 441774/2023-7, 200013/2024-6, 445150/2024-6 and 408990/2025-2), INCT-IQ through the project (465469/2014-0) and National Institute of Science and Technology for Applied Quantum Computing through CNPq process No. 408884/2024-0. 
F.C.B. acknowledges support from Taighde Éireann - Research Ireland under grant number IRCLA/2022/3922.

\end{acknowledgments}

\bibliography{refs.bib}

@article{Brandao2015c,
archivePrefix = {arXiv},
arxivId = {1310.8640},
author = {Brand{\~{a}}o, Fernando G.S.L. and Piani, Marco and Horodecki, Pawe{\l}},
doi = {10.1038/ncomms8908},
eprint = {1310.8640},
issn = {20411723},
journal = {Nature Communications},
pages = {7908},
title = {{Generic emergence of classical features in quantum Darwinism}},
volume = {6},
year = {2015}
}

@article{Zwolak2013,
archivePrefix = {arXiv},
arxivId = {1303.4659},
author = {Zwolak, Michael and Zurek, Wojciech H.},
doi = {10.1038/srep01729},
eprint = {1303.4659},
issn = {20452322},
journal = {Scientific Reports},
pages = {1729},
title = {{Complementarity of quantum discord and classically accessible information}},
volume = {3},
year = {2013}
}

@article{Zurek2009,
archivePrefix = {arXiv},
arxivId = {0903.5082},
author = {Zurek, Wojciech Hubert},
doi = {10.1038/nphys1202},
eprint = {0903.5082},
issn = {17452481},
journal = {Nature Physics},
number = {3},
pages = {181--188},
publisher = {Nature Publishing Group},
title = {{Quantum Darwinism}},
volume = {5},
year = {2009}
}

@article{Modi2012,
archivePrefix = {arXiv},
arxivId = {1112.6238},
author = {Modi, Kavan and Brodutch, Aharon and Cable, Hugo and Paterek, Tomasz and Vedral, Vlatko},
doi = {10.1103/RevModPhys.84.1655},
eprint = {1112.6238},
issn = {00346861},
journal = {Reviews of Modern Physics},
number = {4},
pages = {1655},
title = {{The classical-quantum boundary for correlations: Discord and related measures}},
volume = {84},
year = {2012}
}

@article{latune2025thermodynamically,
  title={A thermodynamically consistent approach to the energy costs of quantum measurements},
  author={Latune, Camille L and Elouard, Cyril},
  journal={Quantum},
  volume={9},
  pages={1614},
  year={2025},
  doi={10.22331/q-2025-03-20-1614}
}

@article{deffner2016quantum,
  title={Quantum work and the thermodynamic cost of quantum measurements},
  author={Deffner, Sebastian and Paz, Juan Pablo and Zurek, Wojciech H},
  journal={Physical Review E},
  volume={94},
  number={1},
  pages={010103},
  year={2016},
  doi={10.1103/PhysRevE.94.010103}
}

@article{Faist2015a,
archivePrefix = {arXiv},
arxivId = {1211.1037},
author = {Faist, Philippe and Dupuis, Fr{\'{e}}d{\'{e}}ric and Oppenheim, Jonathan and Renner, Renato},
doi = {10.1038/ncomms8669},
eprint = {1211.1037},
issn = {20411723},
journal = {Nature Communications},
pages = {7669},
title = {{The minimal work cost of information processing}},
volume = {6},
year = {2015}
}

@article{Sagawa2009,
author = {Sagawa, Takahiro and Ueda, Masahito},
doi = {10.1103/PhysRevLett.102.250602},
journal = {Physical Review Letters},
pages = {250602},
title = {{Minimal Energy Cost for Thermodynamic Information Processing : Measurement and Information Erasure}},
volume = {102},
year = {2009}
}

@article{jacobs2012quantum,
  title={Quantum measurement and the first law of thermodynamics: The energy cost of measurement is the work value of the acquired information},
  author={Jacobs, Kurt},
  journal={Physical Review E},
  volume={86},
  number={4},
  pages={040106},
  year={2012},
  doi={10.1103/PhysRevE.86.040106}
}

@article{linpeng2022energetic,
  title={Energetic cost of measurements using quantum, coherent, and thermal light},
  author={Linpeng, Xiayu and Bresque, L{\'e}a and Maffei, Maria and Jordan, Andrew N and Auff{\`e}ves, Alexia and Murch, Kater W},
  journal={Physical Review Letters},
  volume={128},
  number={22},
  pages={220506},
  year={2022},
  doi={10.1103/PhysRevLett.128.220506}
}

@article{abdelkhalek2016fundamental,
  title={Fundamental energy cost for quantum measurement},
  author={Abdelkhalek, Kais and Nakata, Yoshifumi and Reeb, David},
  journal={arXiv:1609.06981},
URL = {https://arxiv.org/abs/1609.06981},
  year={2016}
}

@article{mohammady2023thermodynamically,
  title={Thermodynamically free quantum measurements},
  author={Mohammady, M Hamed},
  journal={Journal of Physics A: Mathematical and Theoretical},
  volume={55},
  number={50},
  pages={505304},
  year={2023},
  doi={https://doi.org/10.1088/1751-8121/acad4a}
}

@article{elouard2018efficient,
  title={Efficient quantum measurement engines},
  author={Elouard, Cyril and Jordan, Andrew N},
  journal={Physical Review Letters},
  volume={120},
  number={26},
  pages={260601},
  year={2018},
  doi={10.1103/PhysRevLett.120.260601}
}

@article{demelo2024finite,
  title={A finite-resource description of a measurement process and its implications for the "Wigner's Friend" scenario},
  author={de Melo, Fernando and Carvalho, Gabriel Dias and Correia, Pedro S and Obando, Paola Concha and de Oliveira, Thiago R and Vallejos, Ra{\'u}l O},
  journal={arXiv:2411.07327},
URL = {https://arxiv.org/abs/2411.07327},
  year={2024}
}

@article{lipka2024fundamental,
  title={Fundamental limits on anomalous energy flows in correlated quantum systems},
  author={Lipka-Bartosik, Patryk and Diotallevi, Giovanni Francesco and Bakhshinezhad, Pharnam},
  journal={Physical Review Letters},
  volume={132},
  number={14},
  pages={140402},
  year={2024},
  doi={10.1103/PhysRevLett.132.140402}
}

@article{krisnanda2022correlations,
  title={Correlations and energy in mediated dynamics},
  author={Krisnanda, Tanjung and Lee, Su-Yong and Noh, Changsuk and Kim, Jaewan and Streltsov, Alexander and Liew, Timothy CH and Paterek, Tomasz},
  journal={New Journal of Physics},
  volume={24},
  number={12},
  pages={123025},
  year={2022},
  doi={https://doi.org/10.1088/1367-2630/aca9ef}
}

@article{deoliveira2025heat,
  title={Heat as a witness of quantum properties},
  author={de Oliveira Junior, A and Brask, Jonatan Bohr and Lipka-Bartosik, Patryk},
  journal={Physical Review Letters},
  volume={134},
  number={5},
  pages={050401},
  year={2025},
  doi={10.1103/PhysRevLett.134.050401}
}

@article{piccione2021generation,
  title={Generation of minimum-energy entangled states},
  author={Piccione, Nicol{\`o} and Militello, Benedetto and Napoli, Anna and Bellomo, Bruno},
  journal={Physical Review A},
  volume={103},
  number={6},
  pages={062402},
  year={2021},
  doi={10.1103/PhysRevA.103.062402}
}

@article{piccione2020energy,
  title={Energy bounds for entangled states},
  author={Piccione, Nicol{\`o} and Militello, Benedetto and Napoli, Anna and Bellomo, Bruno},
  journal={Physical Review Research},
  volume={2},
  number={2},
  pages={022057},
  year={2020},
  doi={10.1103/PhysRevResearch.2.022057}
}

@article{simon2025correlations,
  title={Correlations enable lossless ergotropy transport},
  author={Simon, Rick PA and Anders, Janet and Hovhannisyan, Karen V},
  journal={Physical Review Letters},
  volume={134},
  number={1},
  pages={010408},
  year={2025},
  doi={10.1103/PhysRevLett.134.010408}
}

@article{kammerlander2016coherence,
  title={Coherence and measurement in quantum thermodynamics},
  author={Kammerlander, Philipp and Anders, Janet},
  journal={Scientific Reports},
  volume={6},
  number={1},
  pages={22174},
  year={2016},
  doi={10.1038/srep22174}
}

@article{engineer2024equilibration,
  title={Equilibration of objective observables in a dynamical model of quantum measurements},
  author={Engineer, Sophie and Rivlin, Tom and Wollmann, Sabine and Malik, Mehul and Lock, Maximilian PE},
  journal={arXiv:2403.18016},
URL = {https://arxiv.org/abs/2403.18016},
  year={2024}
}

@article{Janzing2000,
  author={Janzing, D. and Wocjan, P. and Zeier, R. and Geiss, R. and Beth, Th},
  title={Thermodynamic cost of reliability and low temperatures: Tightening Landauer's principle and the second law},
  journal={International Journal of Theoretical Physics},
  volume={39},
  number={12},
  pages={2717--2753},
  year={2000},
  doi={10.1023/A:1026422630734},
  arxiv={quant-ph/0002048}
}

@article{Brandao2013,
  author={Brand{\~{a}}o, Fernando G.S.L. and Horodecki, Micha{\l} and Oppenheim, Jonathan and Renes, Joseph M and Spekkens, Robert W},
  title={Resource theory of quantum states out of thermal equilibrium},
  journal={Physical Review Letters},
  volume={111},
  number={25},
  pages={250404},
  year={2013},
  doi={10.1103/PhysRevLett.111.250404},
  arxiv={arXiv:1111.3882}
}

@article{GuryanovaFriisHuber2018,
  author={Guryanova, Yelena and Friis, Nicolai and Huber, Marcus},
  title={Ideal Projective Measurements Have Infinite Resource Costs},
  journal={Quantum},
  volume={4},
  pages={222},
  year={2020},
  doi={10.22331/q-2020-01-13-222},
  arxiv={1805.11899}
}

@article{Zurek1991,
  author={Zurek, Wojciech H.},
  title={Decoherence and the transition from quantum to classical},
  journal={Physics Today},
  pages={36--44},
  year={1991},
  doi={10.1063/1.881293}
}

@article{Zurek2002,
  author={Zurek, Wojciech Hubert},
  title={Decoherence and the transition from quantum to classical - Revisited},
  journal={Los Alamos Science},
  volume={27},
  pages={2--25},
  year={2002},
  arxiv={quant-ph/0306072}
}

@article{lostaglio2019introductory,
  title={An introductory review of the resource theory approach to thermodynamics},
  author={Lostaglio, Matteo},
  journal={Reports on Progress in Physics},
  volume={82},
  number={11},
  pages={114001},
  year={2019},
  doi={https://doi.org/10.1088/1361-6633/ab46e5}
}

@incollection{ng2019resource,
  title={Resource theory of quantum thermodynamics: Thermal operations and second laws},
  author={Ng, Nelly Huei Ying and Woods, Mischa Prebin},
  booktitle={Thermodynamics in the quantum regime: Fundamental aspects and new directions},
  publisher={Springer},
  pages={625--650},
  year={2019},
  doi={10.1007/978-3-319-99046-0_25}
}

@article{mohammady2023quantum,
  title={Quantum measurements constrained by the third law of thermodynamics},
  author={Mohammady, M Hamed and Miyadera, Takayuki},
  journal={Physical Review A},
  volume={107},
  number={2},
  pages={022406},
  year={2023},
  doi={10.1103/PhysRevA.107.022406}
}

@article{mohammady2025thermodynamic,
  title={The thermodynamic trilemma of efficient measurements},
  author={Mohammady, M Hamed and Buscemi, Francesco},
  journal={arXiv:2502.14136},
URL = {https://arxiv.org/abs/2502.14136},
  year={2025}
}

@article{krisut2024deriving,
  title={Deriving Tsallis entropy from non-extensive Hamiltonian within a statistical mechanics framework},
  author={Krisut, Paradon and Yoo-Kong, Sikarin},
  journal={arXiv:2411.16757},
URL = {https://arxiv.org/abs/2411.16757},
  year={2024}
}

@article{tsallis2011nonadditive,
  title={The nonadditive entropy Sq and its applications in physics and elsewhere: Some remarks},
  author={Tsallis, Constantino},
  journal={Entropy},
  volume={13},
  number={10},
  pages={1765--1804},
  year={2011},
  doi={10.3390/e13101765}
}

@article{tsallis1988possible,
  title={Possible generalization of Boltzmann-Gibbs statistics},
  author={Tsallis, Constantino},
  journal={Journal of statistical physics},
  volume={52},
  pages={479--487},
  year={1988},
  doi={10.1007/BF01016429}
}

@article{mironowicz2018system,
  title={System information propagation for composite structures},
  author={Mironowicz, Piotr and Nale{\.z}yty, Pawe{\l} and Horodecki, Pawe{\l} and Korbicz, Jaros{\l}aw K},
  journal={Physical Review A},
  volume={98},
  number={2},
  pages={022124},
  year={2018},
  doi={10.1103/PhysRevA.98.022124}
}

@article{blume2006quantum,
  title={Quantum Darwinism: Entanglement, branches, and the emergent classicality of redundantly stored quantum information},
  author={Blume-Kohout, Robin and Zurek, Wojciech H},
  journal={Physical Review A},
  volume={73},
  number={6},
  pages={062310},
  year={2006},
  doi={10.1103/PhysRevA.73.062310}
}

@article{PhysRevA.109.032221,
  title={Complementarity between decoherence and information retrieval from the environment},
  author={Lee, Tae-Hun and Korbicz, Jaros\l{}aw K.},
  journal={Physical Review A},
  volume={109},
  number={3},
  pages={032221},
  year={2024},
  doi={10.1103/PhysRevA.109.032221}
}

@article{lee2024encoding,
  title={Encoding position by spins: Objectivity in the boson-spin model},
  author={Lee, Tae-Hun and Korbicz, Jaros{\l}aw K},
  journal={Physical Review A},
  volume={109},
  number={5},
  pages={052204},
  year={2024},
  doi={10.1103/PhysRevA.109.052204}
}

@article{panda2023nonideal,
  title={Nonideal measurement heat engines},
  author={Panda, Abhisek and Binder, Felix C and Vinjanampathy, Sai},
  journal={Physical Review A},
  volume={108},
  number={6},
  pages={062214},
  year={2023},
  doi={10.1103/PhysRevA.108.062214},
  arxiv={2308.02381}
}

@article{debarba2019work,
  title={Work estimation and work fluctuations in the presence of non-ideal measurements},
  author={Debarba, Tiago and Manzano, Gonzalo and Guryanova, Yelena and Huber, Marcus and Friis, Nicolai},
  journal={New Journal of Physics},
  volume={21},
  number={11},
  pages={113002},
  year={2019},
  doi={10.1088/1367-2630/ab4d9d},
  arxiv={1902.08568}
}

@article{xuereb2024resources,
  title={What resources do agents need to acquire knowledge in Quantum Thermodynamics?},
  author={Xuereb, Jake and Junior, A and Clivaz, Fabien and Bakhshinezhad, Pharnam and Huber, Marcus},
  journal={arXiv:2410.18167},
URL = {https://arxiv.org/abs/2410.18167},
  year={2024}
}

@article{zwolak2010redundant,
  title={Redundant imprinting of information in nonideal environments: Objective reality via a noisy channel},
  author={Zwolak, Michael and Quan, HT and Zurek, Wojciech H},
  journal={Physical Review A},
  volume={81},
  number={6},
  pages={062110},
  year={2010},
  doi={10.1103/PhysRevA.81.062110}
}

@article{paris2013asymptotics,
  title={Asymptotics of the Gauss hypergeometric function with large parameters, I},
  author={Paris, Richard B},
  journal={Journal of Classical Analysis},
  volume={2},
  number={2},
  pages={183--203},
  year={2013},
  doi={10.7153/jca-02-11}
}

@article{temme2003large,
  title={Large parameter cases of the Gauss hypergeometric function},
  author={Temme, Nico M},
  journal={Journal of Computational and Applied Mathematics},
  volume={153},
  number={1-2},
  pages={441--462},
  year={2003},
  doi={https://doi.org/10.1016/S0377-0427(02)00627-1}
}

@article{cvitkovic2017asymptotic,
  title={Asymptotic expansions of the hypergeometric function with two large parameters—application to the partition function of a lattice gas in a field of traps},
  author={Cvitkovi{\'c}, Mislav and Smith, Ana-Sun{\v{c}}ana and Pande, Jayant},
  journal={Journal of Physics A: Mathematical and Theoretical},
  volume={50},
  number={26},
  pages={265206},
  year={2017},
  doi={https://doi.org/10.1088/1751-8121/aa7213}
}

@article{schlosshauer2019quantum,
  title={Quantum decoherence},
  author={Schlosshauer, Maximilian},
  journal={Physics Reports},
  volume={831},
  pages={1--57},
  year={2019},
  doi={10.1016/j.physrep.2019.10.001}
}

@article{zurek2022quantum,
  title={Quantum theory of the classical: Einselection, envariance, quantum darwinism and extantons},
  author={Zurek, Wojciech Hubert},
  journal={Entropy},
  volume={24},
  number={11},
  pages={1520},
  year={2022},
  doi={10.3390/e24111520}
}

@article{schlosshauer2004decoherence,
  title={Decoherence, the measurement problem, and interpretations of quantum mechanics},
  author={Schlosshauer, Maximilian},
  journal={Reviews of Modern Physics},
  volume={76},
  number={4},
  pages={1267--1305},
  year={2004},
  doi={10.1103/RevModPhys.76.1267}
}

@article{zurek2003decoherence,
  title={Decoherence, einselection, and the quantum origins of the classical},
  author={Zurek, Wojciech Hubert},
  journal={Reviews of Modern Physics},
  volume={75},
  number={3},
  pages={715},
  year={2003},
  doi={10.1103/RevModPhys.75.715}
}

@article{le2019strong,
  title={Strong quantum darwinism and strong independence are equivalent to spectrum broadcast structure},
  author={Le, Thao P and Olaya-Castro, Alexandra},
  journal={Physical Review Letters},
  volume={122},
  number={1},
  pages={010403},
  year={2019},
  doi={10.1103/PhysRevLett.122.010403}
}

@article{Feller2021,
  title = {Comment on ``Strong Quantum Darwinism and Strong Independence Are Equivalent to Spectrum Broadcast Structure''},
  author = {Feller, Alexandre and Roussel, Benjamin and Fr\'erot, Ir\'en\'ee and Degiovanni, Pascal},
  journal = {Physical Review Letters},
  volume = {126},
  issue = {18},
  pages = {188901},
  numpages = {2},
  year = {2021},
  month = {May},
  publisher = {American Physical Society},
  doi = {10.1103/PhysRevLett.126.188901},
  url = {https://link.aps.org/doi/10.1103/PhysRevLett.126.188901}
}

@article{Le2021,
  title = {Le and Olaya-Castro Reply:},
  author = {Le, Thao P. and Olaya-Castro, Alexandra},
  journal = {Physical Review Letters},
  volume = {126},
  issue = {18},
  pages = {188902},
  numpages = {1},
  year = {2021},
  month = {May},
  publisher = {American Physical Society},
  doi = {10.1103/PhysRevLett.126.188902},
  url = {https://link.aps.org/doi/10.1103/PhysRevLett.126.188902}
}

@article{horodecki2015quantum,
  title={Quantum origins of objectivity},
  author={Horodecki, Ryszard and Korbicz, JK and Horodecki, Pawe{\l}},
  journal={Physical Review A},
  volume={91},
  number={3},
  pages={032122},
  year={2015},
  doi={10.1103/PhysRevA.91.032122}
}

@article{korbicz2021roads,
  title={Roads to objectivity: quantum darwinism, spectrum broadcast structures, and strong quantum darwinism--a review},
  author={Korbicz, Jarek K},
  journal={Quantum},
  volume={5},
  pages={571},
  year={2021},
  doi={https://doi.org/10.22331/q-2021-11-08-571}
}

@article{mironowicz2017monitoring,
  title={Monitoring of the process of system information broadcasting in time},
  author={Mironowicz, Piotr and Korbicz, JK and Horodecki, Pawe{\l}},
  journal={Physical Review Letters},
  volume={118},
  number={15},
  pages={150501},
  year={2017},
  doi={10.1103/PhysRevLett.118.150501}
}

@article{bae2015quantum,
  title={Quantum state discrimination and its applications},
  author={Bae, Joonwoo and Kwek, Leong-Chuan},
  journal={Journal of Physics A: Mathematical and Theoretical},
  volume={48},
  number={8},
  pages={083001},
  year={2015},
  doi={10.1088/1751-8113/48/8/083001}
}

@article{bergou2010discrimination,
  title={Discrimination of quantum states},
  author={Bergou, J{\'a}nos A},
  journal={Journal of Modern Optics},
  volume={57},
  number={3},
  pages={160--180},
  year={2010},
  doi={10.1080/09500340903477756}
}

@article{zwolak2009quantum,
  title={Quantum Darwinism in a mixed environment},
  author={Zwolak, Michael and Quan, HT and Zurek, Wojciech H},
  journal={Physical Review Letters},
  volume={103},
  number={11},
  pages={110402},
  year={2009},
  doi={10.1103/PhysRevLett.103.110402}
}

@article{TarantoBakhshinezhadEtAl2023,
  author={Taranto, Philip and Bakhshinezhad, Faraj and Bluhm, Andreas and Silva, Ralph and Friis, Nicolai and Lock, Maximilian P. E. and Vitagliano, Giuseppe and Binder, Felix C. and Debarba, Tiago and Schwarzhans, Emanuel and Clivaz, Fabien and Huber, Marcus},
  title={Landauer Versus Nernst: What is the True Cost of Cooling a Quantum System?},
  journal={PRX Quantum},
  volume={4},
  pages={010332},
  year={2023},
  doi={10.1103/PRXQuantum.4.010332},
  arxiv={2106.05151}
}

@book{Joos2003,
author = {Joos, E. and Zeh, H.D. and Kiefer, C. and Giulini, D. and Stamatescu, I.-O. and Kupsch, J.},
doi = {10.1007/978-3-662-05328-7},
edition = {second},
isbn = {978-3-642-05576-8},
publisher = {Springer Berlin / Heidelberg},
title = {{Decoherence and the Appearance of a Classical World in Quantum Theory}},
year = {2003}
}

@book{Schlosshauer2007,
  author    = {Maximilian Schlosshauer},
  title     = {Decoherence and the Quantum-to-Classical Transition},
  publisher = {Springer},
  year      = {2007},
  series    = {The Frontiers Collection},
  isbn      = {978-3-540-35774-1},
  doi       = {10.1007/978-3-540-35775-8},
  address   = {Berlin, Heidelberg}
}

@article{bruschi2015thermodynamics,
  title={Thermodynamics of creating correlations: Limitations and optimal protocols},
  author={Bruschi, David Edward and Perarnau-Lobet, Mart{\'\i} and Friis, Nicolai and Hovhannisyan, Karen V and Huber, Marcus},
  journal={Physical Review E},
  volume={91},
  number={3},
  pages={032118},
  year={2015},
  doi={10.1103/PhysRevE.91.032118}
}

@misc{NIST:DLMF,
  key={DLMF},
  title={NIST Digital Library of Mathematical Functions},
  howpublished={\url{https://dlmf.nist.gov/}, Release 1.2.2 of 2024-09-15},
  note={F.~W.~J. Olver, A.~B. {Olde Daalhuis}, D.~W. Lozier, B.~I. Schneider, R.~F. Boisvert, C.~W. Clark, B.~R. Miller, B.~V. Saunders, H.~S. Cohl, and M.~A. McClain, eds.}
}

@article{touilBranchingStatesEmergent2024,
  title={Branching States as The Emergent Structure of a Quantum Universe},
  author={Touil, Akram and Anza, Fabio and Deffner, Sebastian and Crutchfield, James P.},
  journal={Quantum},
  volume={8},
  pages={1494},
  year={2024},
  doi={10.22331/q-2024-10-10-1494}
}

@article{huber2015thermodynamic,
  title={Thermodynamic cost of creating correlations},
  author={Huber, Marcus and Perarnau-Lobet, Mart{\'\i} and Hovhannisyan, Karen V and Skrzypczyk, Paul and Kl{\"o}ckl, Claude and Brunner, Nicolas and Ac{\'\i}n, Antonio},
  journal={New Journal of Physics},
  volume={17},
  number={6},
  pages={065008},
  year={2015},
  doi={10.1088/1367-2630/17/6/065008}
}

@article{bakhshinezhad2019thermodynamically,
  title={Thermodynamically optimal creation of correlations},
  author={Bakhshinezhad, Faraj and Clivaz, Fabien and Vitagliano, Giuseppe and Erker, Paul and Rezakhani, Ali and Huber, Marcus and Friis, Nicolai},
  journal={Journal of Physics A: Mathematical and Theoretical},
  volume={52},
  number={46},
  pages={465303},
  year={2019},
  doi={https://doi.org/10.1088/1751-8121/ab3932}
}

@book{busch2016quantum,
  title={Quantum measurement},
  author={Busch, Paul and Lahti, Pekka and Pellonp{\"a}{\"a}, Juha-Pekka and Ylinen, Kari},
  volume={23},
  year={2016},
  publisher={Springer},
  doi={10.1007/978-3-319-43389-9}
}

@article{le2021thermality,
  title={Thermality versus objectivity: Can they peacefully coexist?},
  author={Le, Thao P and Winter, Andreas and Adesso, Gerardo},
  journal={Entropy},
  volume={23},
  number={11},
  pages={1506},
  year={2021},
  doi={10.3390/e23111506}
}

@article{allahverdyan2013understanding,
  title={Understanding quantum measurement from the solution of dynamical models},
  author={Allahverdyan, Armen E and Balian, Roger and Nieuwenhuizen, Theo M},
  journal={Physics Reports},
  volume={525},
  number={1},
  pages={1--166},
  year={2013},
  doi={https://doi.org/10.1016/j.physrep.2012.11.001}
}

@article{debarba2024broadcasting,
  title={Unknown measurement statistics cannot be redundantly copied using finite resources},
  author={Debarba, Tiago and Huber, Marcus and Friis, Nicolai},
  journal={arXiv:2403.07660},
URL = {https://arxiv.org/abs/2403.07660},
  year={2024}
}

@article{schwarzhans2023,
  author={Schwarzhans, Emanuel and Binder, Felix C. and Huber, Marcus and Lock, Maximilian P. E.},
  title={Quantum measurements and equilibration: the emergence of objective reality via entropy maximisation},
  journal={arXiv:2302.11253},
url={https://arxiv.org/abs/2302.11253},
  year={2023}
}

@article{ozawa2019intersubjectivity,
  title={Intersubjectivity of outcomes of quantum measurements},
  author={Ozawa, Masanao},
  journal={arXiv:1911.10893},
URL = {https://arxiv.org/abs/1911.10893v1},
  year={2019}
}

\newpage
\appendix

\begin{widetext}
    \section{Generalisation of Ozawa's theorem}
\label{app:ozawa_th}
We present a generalised proof of Ozawa’s Theorem~\cite{ozawa2019intersubjectivity}, extending the original result from pure states to arbitrary density matrices. For clarity, we first consider the minimal case of two environmental subsystems ($N_P = 2$). The extension to $N_P > 2$ follows naturally through identical reasoning. This generalisation preserves the theorem’s core insight while accommodating the mixed-state scenarios inherent in thermodynamic contexts.

We recall that Ozawa's theorem states that if locality and probability reproducibility are satisfied, then the agreement condition is automatically fulfilled. In this context, locality implies the existence of local measurements $\{\LocM{i}{x}\}_{x=0}^{\dS-1}$ and local probability distributions $\tp\Pii(x_i)$, while probability reproducibility means that each observer's measurement obeys the same probability distribution originally derived from the system $\tp\Pii(x) = \tp\Pij(x) = p\Sys(x)$. We can write the probabilities in the Heisenberg picture as
\begin{align}
    \tp\PiOne(x) = \tr{\mathbb{I}_S\otimes \Pi^{(1)}_{x} \otimes \mathbb{I}_{P_2} \trho\SPi} 
     = \tr{\Pi^{(1)}_{x}(t) \rho\Sys \otimes \rho\PiOne \otimes \rho\PiTwo} \, ,
\end{align}
where $\LocM{1}{x}(t) = U^\dagger (\mathbb{I}_S\otimes \Pi^{(1)}_{x} \otimes \mathbb{I}_{P_2}) U$ is the operator in the Heisenberg picture. Similarly, this must hold also for $\tp\PiTwo(x)$ and $\LocM{2}{x}(t)$. Assuming now probability reproducibility, we have that
\begin{equation}
    \tr{\LocM{i}{x}(t) \rho\Sys \otimes \rho\PiOne \otimes \rho\PiTwo} = \bra{x} \rho\Sys \ket{x} \, ,
     \label{eq:HpictPR}
\end{equation}
for $i=1,2$. The last equation must hold for all initial factorised states, and in particular, for any arbitrary pure state $\ket{\psi^0} = \ket{ \psi^0\Sys} \ket{\psi^0\PiOne}\ket{\psi^0\PiTwo}$. Given that $A^{(S)}_x(0)= \ketbra{x}{x}$, it follows that $\LocM{1}{x}(t)$ and $\LocM{2}{x}(t)$ are all orthogonal projectors and we have that Eq. \eqref{eq:HpictPR} can be written as
\begin{equation}
    A^{(S)}_x(0) \vert \psi_0\rangle = \LocM{1}{x}(t) \vert \psi_0\rangle = \LocM{2}{x}(t) \vert \psi_0\rangle\, .
\end{equation}
This means that, by linearity, it also holds for any separable density matrix $\rho_0 = \sum_{j} q_j \vert \psi_0^{(j)}\rangle\langle \psi_0^{(j)}\vert$ (and thus also for any factorised state), implying
\begin{align}
    \tr{\LocM{1}{x}(t)\LocM{2}{y}(t)\rho_0} & = \delta_{xy} \tr{A^{(S)}_x(0) \rho_0} = \\
    & = \delta_{xy} \langle x \vert \rho_S \vert x \rangle\, ,
\end{align}
which is nothing but the agreement condition for two environments. We have thus proved that locality and probability reproducibility imply agreement even in the case of a mixed initial state, extending Ozawa's result to this scenario.

\section{Proof of Theorem \ref{th:fres-intsub}}
\label{app:proof-fres-intsub}
In this section, we provide the formal proof of our main Theorem~\ref{th:fres-intsub}. The initial state of the system can be written as
\begin{equation}
    \rho\Sys = \sum_{x=0}^{\dS-1} p\Sys(x) \vert x \rangle\langle x \vert + \rho\Sys^{\text{off}} \, ,
\end{equation}
where $\rho\Sys^{\text{off}}$ is the off-diagonal part of the initial system, which is irrelevant to our information broadcasting analysis. The initial state of the environment is the tensor product
\begin{equation}
    \rho\PiN = \tau_\beta^{\otimes \NP}\, ,
\end{equation}
where $\tau_\beta$ is the thermal state at temperature $\beta$. According to the discussion in Sec. \ref{sec:intsubjfires}, we can split it into $\dS$ orthogonal components
\begin{equation}
    \tau_\beta = \sum_{x=0}^{\dS-1} A_x  = \bigoplus_{x=0}^{\dS-1} A_x\, ,
\end{equation}
where we stressed the orthogonality of the $\{A_x\}$ with the last notation, and $A_x = \sum_{i \in D_x} p(E_{k}^{(x)}) \vert E_i \rangle \langle E_i \vert \,$, and with the Boltzmann factor $ p(E_{k}^{(y)}) = \frac{e^{-\beta E_{k}^{(y)}}}{Z_\beta}$. Following this decomposition, we can further decompose the initial state of the environment as
\begin{equation}
    \rho\PiN = \rho^{\text{agr}}\PiN \oplus \rho^{\text{dis}}\PiN\, ,
\end{equation}
where $\rho^{\text{agr}}\PiN$ is the component supported on the agreement subspace $\mathcal{A} = \text{span}\{\otimes_{i=1}^{\NP} \Pi^{(i)}_x\}_{x=0}^{\dS-1}$, and thus can be written as
\begin{equation}
    \rho\PiN^{\text{agr}} = \bigoplus_{x=0}^{\dS-1}(A_x)^{\otimes \NP} \, ,
\end{equation}
while $\rho^{\text{dis}}\PiN$ is the orthogonal component. The initial state of the system and environments is then
\begin{equation}
    \rho\SPiN = \sum_{x=0}^{\dS-1} p\Sys(x) \vert x \rangle \langle x \vert \otimes \rho^{\text{agr}}\PiN + \sum_{x=0}^{\dS-1} p\Sys(x) \vert x \rangle \langle x \vert \otimes \rho^{\text{dis}}\PiN + \rho^{\text{off}}\Sys\otimes \rho\PiN \, .
\end{equation}
The first term 
\begin{equation}
    \rho\SPiN^{\text{agr}} = \rho^{\text{dia}}\Sys \otimes \rho^{\text{agr}}\PiN
\end{equation}
is the relevant part for intersubjectivity, since it is supported on diagonal components of the system $\rho^{\text{dia}}\Sys$ and the agreement subspace $\rho^{\text{agr}}\PiN$. To achieve maximum agreement, it is sufficient to evaluate the optimal broadcast between the system and $\mathcal{A}$, given that sending information to the orthogonal subspace $\mathcal{A}_\perp$ will only reduce the agreement and that the off-diagonal part $\rho^\text{off}\Sys$ of the system does not contain the information to be broadcast.
\par
Since we focus on the evolution within the subspace $\mathcal{H}\Sys\otimes \mathcal{H}_{\mathcal{A}}$, where $\mathcal{H}_{\mathcal{A}}$ is the Hilbert space corresponding to the vector space $\mathcal{A}$, we will not obtain a unique unitary on the whole Hilbert space composed of system and environments. Instead, there exists a class $\UBJB{opt}$ of unitaries that maximizes agreement, and we will provide a specific example of such a unitary later.
Hence, our starting point is the state
\begin{equation}
    \rho\SPiN^{\text{agr}} = \sum_{x=0}^{\dS-1} p\Sys(x) \vert x \rangle \langle x \vert \otimes \bigoplus_{y=0}^{\dS-1} \tilde{A}_y\, ,
    \label{eq:redSPiN}
\end{equation}
where we have defined $\tilde{A}_y = A_y^{\otimes \NP}$. Given that we have to satisfy the condition of joint information broadcasting, we can use Lemma 2 from \cite{GuryanovaFriisHuber2018}, which we restate here for the current context:
\begin{lemma}\label{lem:2Gurya} \cite{GuryanovaFriisHuber2018}
    Let $\mathcal{M}_U(\beta)$ be a finite resource measurement procedure consisting of a unitary correlating step between a system and $\mathcal{H}_P$. If the initial state can be written as 
    \begin{equation}
        \rho\SP = \rho\Sys \otimes \sum_{y=0}^{\dS-1} A_y \, ,
        \label{eq:stateLemma2}
    \end{equation}
    then the unitary map $U$ that realises the correlating step and satisfies unbiasedness can be decomposed as two distinct unitaries $\tilde{U}$ and $V$:
    \begin{align}
        \tilde{U} & = \sum_{x=0}^{\dS-1} \vert x\rangle\langle x\vert \otimes \tilde{U}_x \label{eq:Utilde} \\
        V & = \sum_{x,y}^{\dS-1} \sum_{m} \vert y\rangle \langle x \vert \otimes \vert {\psi}_m^{(x)} \rangle \langle {\psi}_m^{(y)} \vert \,,
    \end{align} 
    with $\tilde{U}_{x}$ arbitary unitary on $\mathcal{H}_P$, while $\{\vert {\psi}^{(x)}_m\rangle\}_m$ span the support of $A_x$ for all $x=0,...,\dS-1$.
\end{lemma}
The theorem is given for the case in which all the $A_{x}$ have the same dimension, but this can be relaxed at the cost of a more complicated $V$.  We note the analogy with our reduced problem of biased joint information broadcasting, in particular the resemblance of Eq. \eqref{eq:redSPiN} and Eq. \eqref{eq:stateLemma2}. Furthermore, the unbiasedness at the level of the lemma can be translated to the maximisation of agreement, i.e., the broadcasting of information on the agreement subspace. Indeed, formally, our problem on $\mathcal{H}\Sys\otimes \mathcal{H}_{\mathcal{A}}$ can be recast exactly in the language of Lemma \ref{lem:2Gurya}. More specifically, we can identify $A_y$ with $\tilde{A}_y$, and then apply the theorem. Even though $\rho^{\text{agr}}\SPiN$ is not a normalized state, the proof of the Lemma above still holds. This means that the optimal unitary on $\mathcal{H}\Sys\otimes \mathcal{H}_{\mathcal{A}}$ that achieve the unbiasedness on the agreement subspace is given by an arbitrary step $\tilde{U}$ as in Eq. \eqref{eq:Utilde} (that does not change the agreement and the bias, and thus will be neglected in the following), and a broadcasting step given by
\begin{equation}
    V = \sum_{x,y=0}^{\dS-1} \sum_m \vert y\rangle \langle x \vert \otimes \vert \tilde{\psi}^{(x)}_m \rangle \langle \tilde{\psi}^{(y)}_m \vert\, ,
\end{equation}
where now $\{\vert \tilde{\psi}^{(i)}_m \rangle\}_m$ span the support of $\tilde{A}_i$. We can further expand on the the $\{\vert \tilde{\psi}^{(i)}_m \rangle\}$. Indeed, we have that
\begin{equation}
    \tilde{A}_y = \bigotimes_{i=1}^{\NP} A_y = \bigotimes_{i=1}^{\NP} \sum_{k_i \in D_y} p(E^{(y)}_{k_i}) \vert E_{k_i} \rangle \langle E_{k_i} \vert = \sum_{k_1,...,k_{\NP} \in D_y} \bigotimes_{i=1}^{\NP} p(E^{(y)}_{k_i}) \vert E_{k_i} \rangle \langle E_{k_i} \vert\,.
\end{equation}
We thus identify $m$ as the multi-indixed vector $(k_1,...,k_{\NP})$ with $k_{i} \in D_y$ and 
\begin{equation}
    \vert \tilde{\psi}^{(y)}_m \rangle = \vert E_{k_1} E_{k_2}... E_{k_{\NP}} \rangle, \quad k_1,...,k_{\NP} \in D_y\, .
\end{equation}
In this way, we can write 
\begin{equation}
    \sum_m \vert \tilde{\psi}^{(x)}_m \rangle \langle \tilde{\psi}^{(y)}_m \vert = \sum_{k_1} \vert E_{k_1}^{(x)}\rangle \langle E_{k_1}^{(y)} \vert \otimes ... \otimes \sum_{k_{\NP}} \vert E_{k_{\NP}}^{(x)} \rangle \langle E_{k_{\NP}}^{(y)} \vert= T_{x,y} \otimes ... \otimes T_{x,y} = T_{x,y}^{\otimes \NP} \, ,
\end{equation}
where we have defined $T_{x,y}$ as the transition between the subpsace $y$ to subspace $x$ on a single environment, and set $\vert E_{k}^{(y)}\rangle$ equivalent to $\vert E_{k}\rangle$ for $k\in D_y$. This leads to an operator in the form of
\begin{equation}
    V = \sum_{x,y=0}^{\dS-1} \vert y\rangle \langle x \vert \otimes T_{x,y}^{\otimes \NP}\, .
    \label{eq:app:Vunitary}
\end{equation}
To ensure that $V$ is a unitary operator, it must be extended to act on $\mathcal{H}\Sys \otimes \mathcal{H}_{\mathcal{D}}$, where some degree of arbitrariness exists in the choice of this extension. This could, in principle, affect the bias, but not the agreement, which is evaluated only on the subspace where $V$ acts. Thus, we postpone the discussion of the extension $V_{\mathcal{D}}$ until we address the bias in App. \ref{app:proof-fres-biasBJIB}.
\par
We are now ready to evaluate the maximum agreement. The post-interaction state on the agreement subspace only is
\begin{align}
    \trho\SPiN^{\text{agr}} &= V \left( \sum_{x=0}^{\dS-1} p\Sys(x) \vert x \rangle \langle x \vert \otimes \sum_{y=0}^{\dS-1} (A_y)^{\otimes \NP} \right)V^\dagger = \\
    &= \left(\sum_{x',y'} \ketbra{y'}{x'}\otimes T_{x',y'}^{\otimes \NP}\right)
    \left(\sum_{x=0}^{n-1} p\Sys(x)\ketbra{x}{x}\otimes\sum_{y=0}^{\dS-1} (A_y)^{\otimes \NP}\right) \left(\sum_{x'',y''} \ketbra{x''}{y''}\otimes T_{y'',x''}^{\otimes \NP} \right)\\
    &= \sum_{x,x',x''} \sum_{y, y',y''} \delta_{x,x'}\delta_{x,x''}p\Sys(x) \ketbra{y'}{y''}\otimes \left(T_{x',y'}A_y T_{y'',x''}\right)^{\otimes \NP}.
\end{align}
However, as the subspaces are orthogonal, we have  
\begin{align}
    T_{x',y'}A_y T_{y'',x''} & = \sum_{k}  \vert E_{k}^{(x')} \rangle \langle E_{k}^{(y')} \vert \sum_{k'} p(E_{k'}^{(y)}) \vert E_{k'}^{(y)} \rangle \langle E_{k'}^{(y)} \vert \sum_{k''}  \vert E_{k''}^{(y'')}\rangle \langle E_{k''}^{(x'')} \vert = \\
    & = \delta_{y',y} \delta_{y'',y} \sum_{k} p(E_{k}^{(y)}) \vert E_{k}^{(x')} \rangle \langle E_{k}^{(x'')} \vert \, .
\end{align}
Using this expression in the state evolution and defining $A_{y,x} = \sum_{k} p(E^{(y)}_k) \vert E_k^{(x)} \rangle \langle E_k^{(x)} \vert$, we get
\begin{align}
    \trho\SPiN^{\text{agr}} & =  \sum_{x,x',x''} \sum_{y, y',y''} \delta_{x,x'}\delta_{x,x''} \delta_{y',y} \delta_{y'',y} p\Sys(x) \ketbra{y'}{y''} \otimes \left(\sum_{k} p(E_{k}^{(y)}) \vert E_{k}^{(x')}\rangle \langle  E_{k}^{(x'')} \vert\right)^{\otimes \NP} = \\
    & = \sum_{x,y} p\Sys(x) \ketbra{y}{y}\otimes A_{y,x}^{\otimes \NP} \, .
\end{align}
As the matrices $A_{y,x}$ have orthogonal support to each other,
\begin{equation}
    \tr{\Pi_z A_{y,x}} = \delta_{z,x} \ax{y}, 
\end{equation}
with $\ax{y} = \tr{A_{y,x}} = \tr{A_y}$, we can calculate the agreement $\Agr{\UBJB{opt}}$ (where we have identified $\UBJB{opt}$ as the class of evolutions that maximise the agreement as described above) as follows
\begin{align}
    \Agr{\UBJB{opt}} &= \sum_z \tr{\id\Sys\otimes\Pi_z^{\otimes \NP} \trho\SPiN} = \sum_z \tr{\id\Sys\otimes\Pi_z^{\otimes \NP} \trho\SPiN^{\text{agr}}} = \\ 
    &= \sum_z \sum_{x,y}p\Sys(x) \tr{\Pi_z^{\otimes \NP} A_{y,x}^{\otimes \NP}} \\
    &= \sum_z \sum_{x,y}p\Sys(x) \tr{\Pi_z A_{y,x}}^{\NP} \\
    &= \sum_z \sum_{x,y}p\Sys(x) \delta_{z,x} \ax{y}^{\NP}\\
    & = \sum_y \ax{y}^{\NP} \equiv \fagr{\NP}\,.
\end{align}
This proofs Theorem \ref{th:fres-intsub}, since any other unitary $U$ will have smaller agreement.

\section{Proof of Theorem \ref{th:fres-biasBJIB}}
\label{app:proof-fres-biasBJIB}
In this appendix, we prove Theorem \ref{th:fres-biasBJIB}, which regards the Bias for the optimal unitary $\UBJB{opt}$ discussed in the previous appendix.
\par 
As already discussed, the optimal unitary represents a class of unitaries, since the extension of $V$ to the whole Hilbert space is not uniquely determined by the maximisation of agreement. However, as we will show here, most of the arbitrariness can be eliminated by imposing the condition of BJIB given in Eq. \eqref{eq:bjb}. The only arbitrariness left is the one in $\tilde{U}$, which does not affect any property we are looking at.
\par
The extension of the unitary $V$ given in Eq.~\eqref{eq:app:Vunitary} to the entire Hilbert space is achieved by introducing an additional term $W$ that acts on the orthogonal subspace $\mathcal{H}_{\text{S}} \otimes \mathcal{H}_{\mathcal{D}}$, such that the operator $U = V + W$ remains unitary. Since $W$ acts solely on the orthogonal subspace, we have $VW = 0$. Consequently, $W$ must be unitary within the orthogonal subspace alone, and we can study its properties by looking only at how it acts on $\rho\Sys\otimes \rho\PiN^{\text{dis}}$. More in detail, this reads as
\begin{align}
    \rho^{\text{dis}}\PiN & = \bigotimes_{i=1}^{\NP} \sum_{y_i=0}^{\dS-1} A_{y_i} - \sum_{y=0}^{\dS-1} \bigoplus_{i=1}^{\NP} A_y = \sum_{\{y_1,...y_{\NP}\}} \bigotimes_{i=1}^{\NP} A_{y_i}\, ,
\end{align}
where we have introduced the notation $\sum_{\{y_1,...,y_{\NP}\}}$ to denote the sum over the indices $\{y_1, ..., y_{\NP}\}$, excluding the case where $y_1 = y_2 = ... = y_{\NP}$ (that would correspond to the agreement subspace).
\par
Applying the constraint of BJIB given in Eq. \eqref{eq:bjb}, we need a symmetric broadcast of information across all environments. Given the form of the initial state on $\mathcal{H}\Sys \otimes \mathcal{H}_{\mathcal{D}}$, this is not possible. Thus, in order to satisfy the BJIB, we must avoid acting on $\mathcal{H}\Sys \otimes \mathcal{H}_{\mathcal{D}}$ in order to prevent an asymmetric distribution of the information. We thus conclude that 
\begin{equation}
    W = \id\Sys\otimes \id_{\mathcal{H}_{\mathcal{D}}}\, .
\end{equation}
In this way, the global evolution that maximises agreement and satisfies BJIB is 
\begin{equation}
    U = V + W = \sum_{x,y=0}^{\dS-1} \vert y\rangle \langle x \vert \otimes T_{x,y}^{\otimes \NP} + \id\Sys\otimes \id_{\mathcal{H}_{\mathcal{D}}}\, .
\end{equation}
\par
We can now evaluate the bias, completing the proof of Theorem~\ref{th:fres-biasBJIB}. The action of the whole unitary can be written as
\begin{align}
    \trho\SPiN & = U ( \rho\Sys^{\text{dia}} \otimes \rho^{\text{agr}}\PiN + \rho\Sys^{\text{dia}} \otimes \rho^{\text{dis}}\PiN + \rho^{\text{off}}\Sys \otimes \rho\PiN)U^\dagger = \\
    & = U \rho\Sys^{\text{dia}} \otimes \rho^{\text{agr}}\PiN U^\dagger + U \rho\Sys^{\text{dia}} \otimes \rho^{\text{dis}}\PiN U^\dagger + U \rho^{\text{off}}\Sys \otimes \rho\PiN U^\dagger = \\
    & = V \rho\Sys^{\text{dia}} \otimes \rho^{\text{agr}}\PiN V^\dagger + W \rho\Sys^{\text{dia}} \otimes \rho^{\text{dis}}\PiN W^\dagger + U \rho^{\text{off}}\Sys \otimes \rho\PiN  U^\dagger = \\
    & = \trho\SPiN^{\text{agr}} + \rho\Sys^{\text{dia}} \otimes \rho^{\text{dis}}\PiN + \trho^{\text{off}}\SPiN \,,
\end{align}
where we have used the fact that $V$ acts only on $\rho\Sys^{\text{dia}} \otimes \rho^{\text{agr}}\PiN$ while $W$ acts only on $\rho\Sys^{\text{dia}} \otimes \rho^{\text{dis}}\PiN$. Moreover, $\trho^{\text{off}}\SPiN$ is irrelevant since it is off-diagonal. Evaluating the local probability yields (given the symmetry, the environment to be measured is irrelevant)
\begin{equation}
    \tp\Pii(z) = \tr{\id\Sys \otimes \Pi_z \otimes \id^{\otimes \NP-1}\Poi\trho\SPiN} = \tr{\id\Sys \otimes \Pi_z \otimes \id^{\otimes \NP-1}\Poi\trho\SPiN^{\text{agr}}} + \tr{\id\Sys \otimes \Pi_z \otimes \id^{\otimes \NP-1}\Poi\rho\Sys^{\text{dia}} \otimes \rho^{\text{dis}}\PiN} \, .
\end{equation}
The first term yields
\begin{align}
    \tr{\id\Sys \otimes \Pi_z \otimes \id^{\otimes \NP-1}\Poi\trho\SPiN^{\text{agr}}} & = \tr{\id\Sys \otimes \Pi_z \otimes \id^{\otimes \NP-1}\Poi\sum_{x,y} p\Sys(x) \ketbra{y}{y}\otimes A_{y,x}^{\otimes \NP}} = \\
    & = \sum_{x,y} p\Sys(x) \tr{\Pi_z A_{y,x}} \tr{A_{y,x}}^{\NP-1}\\
    & = \sum_{x,y} p\Sys(x) \delta_{z,x}\ax{y}  \ax{y}^{\NP-1}\\
    & = p\Sys(z) \fagr{\NP}\, ,
\end{align}
while the second yields
\begin{align}
    \tr{\id\Sys \otimes \Pi_z \otimes \id^{\otimes \NP-1}\Poi\rho\Sys^{\text{dia}} \otimes \rho^{\text{dis}}\PiN} & = \tr{\id\Sys \otimes \Pi_z \otimes \id^{\otimes \NP-1}\Poi\sum_x p\Sys(x) \vert x\rangle \langle x \vert \otimes \sum_{\{y_1,...y_{\NP}\}} \bigotimes_{i=1}^{\NP} A_{y_i}} = \\
    & = \tr{\Pi_z \otimes \id\Poi^{\otimes \NP-1} \left(\sum_{y_1,...,y_{\NP}} \bigotimes_{i=1}^{\NP} A_{y_i} - \sum_{y} A_y^{\otimes \NP}\right)} = \\
    & = \sum_{y_1,...,y_{\NP}} \tr{\Pi_z A_{y_1}} \prod_{i=2}^{\NP} \tr{A_{y_i}} - \sum_{y} \tr{\Pi_z A_y}\tr{A_y}^{\otimes \NP-1} \\
    & = \sum_{y_1} \delta_{z,y_1} \ax{y_1} \sum_{y_2,...,y_{\NP}}  \prod_{i=2}^{\NP} \ax{y_i} - \sum_{y} \delta_{z,y} \ax{y} \ax{y}^{\NP-1}  \\
    & = \ax{z} \sum_{y_2,...,y_{\NP}}  \prod_{i=2}^{\NP} \ax{y_i} - \sum_{y} \delta_{z,y} \ax{y} \ax{y}^{\NP-1} \, . 
\end{align}
Given that 
\begin{equation}
    \sum_{y_2,...,y_{\NP}}  \prod_{i=2}^{\NP} \ax{y_i} = \sum_{y_2} \ax{y_2} \sum_{y_3,...,y_{\NP}} \prod_{i=3}^{\NP} \ax{y_i} = \sum_{y_4,...,y_{\NP}} \prod_{i=3}^{\NP} \ax{y_i} =... = 1  \, ,  
\end{equation}
we have that
\begin{equation}
    \tr{\id\Sys \otimes \Pi_z \otimes \id^{\otimes \NP-1}\Poi\rho\Sys^{\text{dia}} \otimes \rho^{\text{dis}}\PiN} = \ax{z} - \ax{z}^{\NP} \, , 
\end{equation}
which eventually leads to
\begin{equation}
    \tp\Pii(z) = p\Sys(z) \fagr{\NP} + \ax{z} -\ax{z}^{\NP} \,,
\end{equation}
which can be rewritten as a mixture of probabilities
\begin{equation}
    \tp\Pii(z) = \fagr{\NP} p\Sys(z)  + (1-\fagr{\NP}) \mst{z} \, ,
\end{equation}
where
\begin{equation}
    \mst{z} = \frac{a(z) -a(z)^\NP}{1-\fagr{\NP}}\, .
\end{equation}
We can eventually evaluate the bias
\begin{align}
	\Bias{\UBJB{\text{opt}}} & = D_{\mathrm{T}}(p\Sys(x),\tp(x)) \\
    & = \frac{1}{2} \sum_{x} \vert p\Sys(x) - \fagr{\NP} p\Sys(x) -(1-\fagr{\NP})\mst{x} \vert \\
    & = \fdis{\NP} D_{\mathrm{T}}(p\Sys(x),\mst{x}). 
\end{align}
This completes the proof of Theorem \ref{th:fres-biasBJIB}. 

\section{Exponential scaling of $\gamma^{(\NP)}_{\textbf{a},\lcg}$ }
\label{app:coarse-grain-proof}

Our aim is to demonstrate the exponential approach of $\fagrl{\NP}{\lcg}$ to $1$ in terms of the dimension of the coarse-graining $\lcg$. We recall the definition of $\fagr{\NP}$
\begin{equation}
    \fagr{\NP} = \sum_{x=0}^{\dS-1} \ax{x}^{\NP}\, .
\end{equation}
The coarse-grained version of the bound $\fagrl{\NP}{\lcg}$ means that we have redefined $\{\axl{x}{\lcg}\}_{x=0}^{\dS-1}$ in terms of products of $\ax{x_i}$. In order to increase $\fagrl{\NP}{\lcg}$, we need that $\axl{0}{\lcg}$ grows to $1$ when $\lcg$ increases, while $\axl{1}{\lcg},...,\axl{\dS-1}{\lcg}$ must decrease with $\lcg$, in order to concentrate the largest eigenvalues in the $A_{0}^{\lcg}$ partition. We can understand this strategy as the one that mimics the closest thing to a pure state.
\par 
The $\axl{x}{\lcg}$ must be defined as the sum over $\ax{x_0}^{k_0}...\ax{x_{\dS-1}}^{k_{\dS-1}}$ with $k_0+..+k_{\dS-1} = \lcg$, given that our initial state is given simply as in Eq. \eqref{eq:cg-state}, that we report here again
\begin{equation}
    \rho\Pilcg = \rho{\Poi}^{\otimes \lcg}  = \sum_{x_1,x_2,..x_\lcg} A_{x_1} \otimes ... \otimes A_{x_\lcg} \, .
\end{equation}
The available eigenvalues to be collected in this $\axl{x}{\lcg}$ are given from the following expansion
\begin{align}
    \tr{\rho{\Poi}^{\otimes \lcg}} & = \sum_{x_1,x_2,...,x_\lcg} \prod_{i=1}^{\lcg}\tr{A_{x_i}} = \sum_{x_1,x_2,...,x_\lcg} \prod_{i=1}^{\lcg}\ax{x_i} = \\
    & = \sum_{k_0+...+k_{\dS-1}=\lcg} \binom{\lcg}{k_0,...,k_{\dS-1}} \ax{0}^{k_0}... \ax{\dS-1}^{k_{\dS-1}} \, , 
    \label{eq:rhoPoiLcg}
\end{align}
where in the last step we have used the multinomial theorem, and where 
\begin{equation}
    \binom{\lcg}{k_0,...,k_{\dS-1}} = \frac{\lcg!}{k_0!...k_{\dS-1}!}
\end{equation}
is the multinomial coefficient. We need to use the elements in the sum in Eq. \eqref{eq:rhoPoiLcg} to define the coarse-grained $\axl{x}{\lcg}$. As one can imagine, dealing with such an expression is not easy. We thus analytically study the simplest scenario with $\dS=2$ in App. \ref{app:coarse-grain-proof-ds2}. We provide the method to study the problem numerically for $d_S > 2$ in App. \ref{app:coarse-grain-proof-dslarger}

\subsection{Proof of Theorem \ref{th:idealintcg} (the case of $\dS=2$)}
\label{app:coarse-grain-proof-ds2}

In this case, the expression for the expansion simplifies to
\begin{equation}
    1 = \sum_{k=0}^{\lcg} \binom{\lcg}{k} \ax{0}^{\lcg-k} \ax{1}^k := \axl{\lcg}{0} + \axl{\lcg}{1}\, .
    \label{eq:binom_ax}
\end{equation}
In order to have neater expressions, we consider only odd $\lcg$, but the results that follow are easily generalizable to even $\lcg$. In Eq. \eqref{eq:binom_ax}, we assign the first half to $\axl{0}{\lcg}$ in such a way that
\begin{equation}
    \axl{0}{\lcg} = \sum_{k=0}^{\frac{\lcg-1}{2}} \binom{\lcg}{k} \ax{0}^{\lcg-k} \ax{1}^k \,.
    \label{eq:decds2ax0lcg}
\end{equation}
Now, we can expand this as follows
\begin{align}
    \sum_{k=0}^{\frac{\lcg-1}{2}} \binom{\lcg}{k} \ax{0}^{\lcg-k} \ax{1}^k = 1 - \sum_{k=\frac{\lcg-1}{2}+1}^{\lcg} \binom{\lcg}{k} \ax{0}^{\lcg-k}\ax{1}^{k}\,.
\end{align}
Making the change of variable $k' = k - (\lcg+1)/2$, we have that
\begin{align}
    \sum_{k=0}^{\frac{\lcg-1}{2}} \binom{\lcg}{k} \ax{0}^{\lcg-k} \ax{1}^k & = 1-\sum_{k'=0}^{\frac{\lcg-1}{2}} \binom{\lcg}{k'+\frac{\lcg+1}{2}} \ax{0}^{\lcg-k'-\frac{\lcg+1}{2}} \ax{1}^{k' + \frac{\lcg+1}{2}} = \\
    & = 1- \ax{0}^{\frac{\lcg-1}{2}} \ax{1}^{\frac{\lcg+1}{2}}\sum_{k'=0}^{\frac{\lcg-1}{2}} \binom{\lcg}{k'+\frac{\lcg+1}{2}} \left(\frac{\ax{1}}{\ax{0}}\right)^{k'}\, .
\end{align}
We further have that
\begin{align}
    \binom{\lcg}{k' + \frac{\lcg+1}{2}} & = \frac{\lcg!}{\left(k' + \frac{\lcg+1}{2}\right)! \left(\lcg - k' - \frac{\lcg+1}{2}\right)!} = \\
    & = \frac{\lcg!}{\left(k' + \frac{\lcg+1}{2}\right)! \left(\frac{\lcg-1}{2} - k'\right)!} = \\
    & = \frac{\lcg!}{\left(\frac{\lcg+1}{2}\right)!\left(\frac{\lcg-1}{2}\right)!} \frac{\left(\frac{\lcg+1}{2}\right)!\left(\frac{\lcg-1}{2}\right)!}{\left(\frac{\lcg+1}{2} + k'\right)!\left(\frac{\lcg-1}{2} -k'\right)!} = \\
    & = \binom{\lcg}{\frac{\lcg+1}{2}} \frac{\left(\frac{\lcg-1}{2}\right)!}{\left(\frac{\lcg-1}{2}-k'\right)!} \frac{\left(\frac{\lcg+1}{2}\right)!}{\left(\frac{\lcg+1}{2}+k'\right)!}\, ,
\end{align}
which leads to
\begin{equation}
    \axl{0}{\lcg} = 1- \ax{0}^{\frac{\lcg-1}{2}}\ax{1}^{\frac{\lcg+1}{2}} \binom{\lcg}{\frac{\lcg+1}{2}} \sum_{k'=0}^{\frac{\lcg-1}{2}}\frac{\left(\frac{\lcg-1}{2}\right)!}{\left(\frac{\lcg-1}{2}-k'\right)!} \frac{\left(\frac{\lcg+1}{2}\right)!}{\left(\frac{\lcg+1}{2}+k'\right)!} \left(\frac{\ax{1}}{\ax{0}}\right)^{k'}\,.
\end{equation}
Considering that the hypergeometric function can be written as \cite{NIST:DLMF}
\begin{equation}
    \hypgeo{2}{1}\left(b,-m;c;z\right) = \sum_{n=0}^{m} (-1)^n \binom{m}{n} \frac{(b)_n}{(c)_n} z^n\, ,
\end{equation}
where $m$ is a positive integer, and $\vert z\vert <1$, and $(p)_n$ is the Pochhammer symbol
\begin{equation}
    (p)_n = 
    \begin{cases}
        1 & n=0 \\
        p(p+1)...(p+n-1) & n>0
    \end{cases}\, .
\end{equation}
In our case, we identify $b=1$, $m=(\lcg-1)/2$, $c = (\lcg+3)/2$ and $z = - \ax{1}/\ax{0}$. Since $(1)_n = n!$ and
\begin{equation}
    \left(\frac{\lcg+3}{2}\right)_k = \frac{\left(\frac{\lcg+3}{2}+k-1\right)!}{\left(\frac{\lcg+3}{2}-1\right)!}.
\end{equation}
We eventually get
\begin{align}
    \hypgeo{2}{1}\left(1,\frac{1-\lcg}{2};\frac{\lcg+3}{2};-\frac{\ax{1}}{\ax{0}}\right) & = \sum_{n=0}^{(\lcg-1)/2}(-1)^n \frac{\left(\frac{\lcg-1}{2}\right)!}{ n!\left(\frac{\lcg-1}{2}-n\right)!}  \frac{n!}{\left(\frac{\lcg+3}{2}+k-1\right)!} \left(\frac{\lcg+3}{2}-1\right)! \left(-\frac{\ax{1}}{\ax{0}}\right)^n= \\
    & = \sum_{n=0}^{(\lcg-1)/2}  \frac{\left(\frac{\lcg-1}{2}\right)!}{ \left(\frac{\lcg-1}{2}-n\right)!}  \frac{\left(\frac{\lcg+1}{2}\right)!}{\left(\frac{\lcg+1}{2}+n\right)!}  \left(\frac{\ax{1}}{\ax{0}}\right)^n
\end{align}
from which we have that
\begin{equation}
    \axl0\lcg = 1- \ax{0}^{\frac{\lcg-1}{2}}\ax{1}^{\frac{\lcg+1}{2}} \binom{\lcg}{\frac{\lcg+1}{2}} \hypgeo{2}{1}\left(1,\frac{1-\lcg}{2};\frac{\lcg+3}{2};-\frac{\ax{1}}{\ax{0}}\right)\, .
\end{equation}
Since we are interested in large $\lcg$, thanks to Stirling's expansion, we have that
\begin{equation}
    \binom{\lcg}{\frac{\lcg+1}{2}} \simeq \frac{2^{\lcg-1/2}}{\sqrt{\pi \lcg}}\, . 
\end{equation}
The hypergeometric function defined above has the following asymptotic behaviour \cite{temme2003large,paris2013asymptotics,cvitkovic2017asymptotic}
\begin{equation}
    \hypgeo{2}{1}(b, a + \epsilon \lambda; c+\lambda;z) \sim \frac{1}{(1-\epsilon z)^b}, \quad \vert \lambda \vert \to +\infty \,.
\end{equation}
In our case we identify $b=1$,$a=1/2$, $\epsilon=-1$, $c=3/2$ and we rescale $\lambda \to \lcg/2$. In this way, we get that
\begin{equation}
    \hypgeo{2}{1}\left(1,\frac{1-\lcg}{2};\frac{\lcg+3}{2};-\frac{\ax{1}}{\ax{0}}\right) \sim \frac{1}{1-\frac{\ax{1}}{\ax{0}}} = \frac{\ax{0}}{\ax{0}-\ax{1}} = C(\bm{a})\, .
\end{equation}
As a last step, we consider
\begin{equation}
    \ax{0}^{\frac{\lcg-1}{2}}\ax{1}^{\frac{\lcg+1}{2}} \frac{2^{\lcg-1/2}}{\sqrt{\pi\lcg}} = \ax{1} \left(\ax{0}\ax{1}\right)^{\frac{\lcg-1}{2}}2^{\lcg-1}\frac{\sqrt{2}}{\sqrt{\pi\lcg}} = \ax{1}\left(2\sqrt{\ax{0}\ax{1}}\right)^{\lcg-1}\frac{\sqrt{2}}{\sqrt{\pi\lcg}}\, .
    \label{eq:dec-poly}
\end{equation}
This quantity decreases as $\lcg$ increase if $2\sqrt{\ax{0}\ax{1}} \leq 1$. Without loosing in generality, we can write $\ax{0} = 1/2 + \varepsilon$ and $\ax{1}=1/2-\varepsilon$, with $\varepsilon\in[0,1/2]$.
In this way
\begin{equation}
    2\sqrt{\ax{0}\ax{1}} = 2 \sqrt{\frac{1}{4}-\varepsilon^2} = \sqrt{1-4\varepsilon^2} \leq 1 \,.
\end{equation}
We have equality iff $\varepsilon=0$, but even in this case the overall expression decreases as $\lcg$ thanks to a term proportional to $\lcg^{-1/2}$, see Eq. \eqref{eq:dec-poly}. We thus define the positive quantity
\begin{equation}
    D(\bm{a}) = - \log 2 \sqrt{\ax{0}\ax{1}}\,,
\end{equation}
in such a way that
\begin{equation}
    \ax{0}^{\frac{\lcg-1}{2}}\ax{1}^{\frac{\lcg+1}{2}} \frac{2^{\lcg-1/2}}{\sqrt{\pi\lcg}} = \ax{1} e^{-D(\bm{a})(\lcg-1)} \sqrt{\frac{2}{\pi \lcg}}\,.
\end{equation}
Thus, we conclude that
\begin{equation}
    \axl{0}\lcg \simeq 1- \frac{e^{-D(\bm{a})(\lcg-1)}}{\sqrt{\lcg}}\ax{1}\sqrt{\frac{2}{\pi}}C(\bm{a}) = 1- \frac{e^{-D(\bm{a})(\lcg-1)}}{\sqrt{\lcg}}F(\bm{a})\,,
\end{equation}
where $F(\bm{a})$ is simply a finite constant
\begin{equation}
    F(\bm{a}) = \sqrt{\frac{2}{\pi}} \frac{\ax{1}{\ax{0}}}{\ax{0}-\ax{1}}\,.
\end{equation}
At the same time, we have that
\begin{equation}
    \axl{1}{\lcg} = \frac{e^{-D(\bm{a})(\lcg-1)}}{\sqrt{\lcg}}F(\bm{a})\,.
\end{equation}
Eventually, using the fact that $(1-x)^N \sim 1-Nx+O(x^2)$ for small $x$, we can consider $\fagrl{\NP}{\lcg}$
\begin{align}
    \fagrl{\NP}{\lcg} & = \axl{0}{\lcg}^{\NP} + \axl{1}{\lcg}^{\NP} \simeq \\
    & \simeq 1-\NP \axl{1}{\lcg} + \axl{1}{\lcg}^{\NP} + O(\axl{1}{\lcg}^2) \simeq \\
    & \simeq 1- \NP \frac{e^{-D(\bm{a})(\lcg-1)}}{\sqrt{\lcg}}F(\bm{a}) + O(\axl{1}{\lcg}^2)\,.
\end{align}
As expected, $\fagrl{\NP}{\lcg}$ converges to $1$ exponentially fast with the dimension of the coarse-graining $\lcg$. We also see the interplay between $\NP$ and $\lcg$ here: a larger $\NP$ increases the gap with respect to the ideal $1$, but the effect is merely linear, and the exponential convergence is not affected by the value of $\NP$.

\subsection{The case of $\dS>2$}
\label{app:coarse-grain-proof-dslarger}

When considering $\dS>2$, deriving an explicit analytical expression for the $\axl{x}{\lcg}$ proves to be challenging. For this reason, we provide numerical evidence of such an exponential scaling to $1$, but only for $\axl{0}{\lcg}$. In this case, the latter quantity is defined from the multinomial expansion as
\begin{equation}
    \axl{0}{\lcg} = \sum_{
    \substack{k_0+...+k_{\dS-1}=\lcg \\
    k_{0} \geq k_j \forall j\neq 0}} \binom{\lcg}{k_0,...,k_{\dS-1}} \ax{0}^{k_0}... \ax{\dS-1}^{k_{\dS-1}} \, .
    \label{eq:ax0lcgds}
\end{equation}
This mimics the behaviour of the decomposition that we employed for $\dS=2$ in Eq. \eqref{eq:decds2ax0lcg} but for larger $\dS$.
\par 
We explore numerically the conjecture that even in this case the $\fagrl{\NP}{\lcg}$ approaches $1$ exponentially fast with $\lcg$. To avoid any complications, we only deal with the scaling $1-\axl{\lcg}{0}$, which proves to be sufficient to prove the scaling of $\fagrl{\NP}{\lcg}$ with some additional rescaling in terms of $\NP$, as in the case of $\dS=2$. 
\par 
For this reason, we want to test if $1-\axl{0}{\lcg}$ behave as a function of the form $c_0 e^{c_1 \lcg}$. In order to perform the fit, we linearise the function by taking the log, i.e., by considering $\ln c_0 + c_1 \lcg$. For fixed $\dS$, we evaluate the $\ln(1-\axl{0}{\lcg})$ using Eq. \eqref{eq:ax0lcgds} for different values of $\{\lcg^{(i)}\}_i$, obtaining a vector of data $\{l_i\}_i$. We then use this data to fit the function and extrapolate $\ln c_0$ and $c_1$. We assess the precision of the fit with the coefficient of determination $R^2$. If $\{f_i\}_i$ are the values that the fitted function takes at the points $\lcg^{(i)}$, we define the residuals as $r_i = l_i-f_i$. Then, $R^2$ is 
\begin{equation}
    R^2 = 1- \frac{S_{\text{res}}}{S_{\text{tot}}}\,,
\end{equation}
where
\begin{align}
    S_{\text{res}} & = \sum_i r_i^2 \\
    S_{\text{tot}} & = \sum_i (l_i -\overline{l})^2\,,
\end{align}
and where $\overline{l}$ is the average of $\{l_i\}_i$. Results are reported in Fig. \ref{fig:ax0lcg} in the main text, with the fit parameters in the table below the figure.

\section{Details on pure dephasing Hamiltonian}
\label{app:ev-puredephH}

In this section, we provide the methods for the simulation of a  star-shaped quantum system specified by the Hamiltonian
\begin{equation}
    \HPD = \frac{g}{2} \sum_{i=1}^{\NP} \sigma_z^{(S)} \otimes \sigma_z^{(i)}\, . 
\end{equation}
To efficiently simulate this system for large $\NP$, we adopt the approach developed in Ref. \cite{zwolak2010redundant}, which utilizes the symmetry of the interaction. In addition to their derivation, here we provide the method we used to efficiently evaluate the optimal discrimination measurement between the two branches, exploiting the block diagonal form of the density matrix.
\par 
Considering an initial product state $\rho(0)= \rho\Sys\otimes_i \rho\Pii$, the evolution is given by $U(t) = \exp(-i \HPD t)$, which can be expanded as a conditioned unitary
\begin{equation}
    U(t) = \vert 0 \rangle \langle 0 \vert \otimes [V(t)^{\otimes \NP}] + \vert 1\rangle\langle 1\vert \otimes [V(-t)^{\otimes \NP}]\, ,
\end{equation}
where $V(t)$ is a unitary matrix given as $V(t) = \exp(-itg \sigma_z /2)$. We are mainly interested in the evolution of the environments, which can be easily written as
\begin{equation}
    \rho\PiN(t) = p_0 \tilde\rho_0(t)^{\otimes \NP} + p_1 \tilde\rho_1(t)^{\otimes \NP}\, ,
\end{equation}
where
\begin{align}
    \tilde\rho_0(t) & = V(t) \rho_{\Poi} V(t)^\dagger \\
    \tilde\rho_1(t) & = V(-t) \rho_{\Poi} V(-t)^\dagger
\end{align}
is the evolution of a single environment state, while $\rho_{\Poi}$ corresponds to its initial state. Further, the state of a coarse-grained macrofraction has the same structure, i.e., for any coarse-graining of size $\lcg$ we have that
\begin{equation}
    \rho\Pilcg(t) = p_0 \tilde{\rho}_0(t)^{\otimes \lcg} + p_1 \tilde{\rho}_1(t)^{\otimes \lcg}\, .
\end{equation}
The strategy for evaluating such an evolution involves rewriting the two branch states, $\tilde{\rho}_x(t)^{\otimes N}$, $x=0,1$, as direct sums of total spin states. This reformulation allows the matrix to be expressed in a block-diagonal form, significantly reducing the memory cost of the computation. Specifically, the largest block to handle has dimensions $(2N+1) \times (2N+1)$, compared to the full density matrix of the same fraction, which scales as $2^N \times 2^N$. This simplification is crucial in our analysis, as it enables us to efficiently determine eigenvalues and the optimal discrimination measurement, and thus investigate how agreement and bias scale with the increasing size of the macrofraction.
\par 
The first step to write $\rho\Poilcg(t)$ in block diagonal form consists of diagonalising the local density matrix $\tilde{\rho}_x(t)$. The diagonalisation can be understood as a rotation of the two-level system, to change the representation from $\sigma_z$ to $\sigma_{\vec{n}}$, where $\vec{n}$ is the vector in the Bloch sphere along which the density matrix is diagonal. The rotation can be written as
\begin{equation}
    R(\alpha,\beta,\gamma) = e^{-i \alpha J_z} e^{-i\beta J_y} e^{-i\gamma J_z}\,,
\end{equation}
where $J_y$ and $J_z$ are the components of the angular momentum in the fundamental representation, i.e., proportional to the Pauli matrices. Therefore, we have that
\begin{equation}
    R(\alpha,\beta,\gamma) \tilde{\rho}_0(t)  R(\alpha,\beta,\gamma)^\dagger = \textup{Diag}[\lambda_+,\lambda_-]\, ,
\end{equation}
and similarly for the other branch
\begin{equation}
    R(-\alpha,-\beta,-\gamma) \tilde{\rho}_1(t)  R(-\alpha,-\beta,-\gamma)^\dagger = \textup{Diag}[\lambda_-,\lambda_+]\,.
\end{equation}
Following \cite{zwolak2010redundant}, the rotation angles can be written as
\begin{align}
    \alpha = \gamma & = t \\
    \sin(\beta/2) & =-\frac{\rho_{00} -\lambda_+}{\sqrt{\vert \rho_{01}\vert^2 + (\rho_{00}-\lambda_+)^2}} \\
    \cos(\beta/2) & = \frac{\rho_{01}}{\sqrt{\vert\rho_{01}\vert^2 + (\rho_{00}-\lambda_+)^2}}\,,
\end{align}
where $\rho_{00} = \langle 0\vert \rho\Poi \vert 0\rangle$ and $\rho_{01} = \langle 0\vert \rho\Poi \vert 1\rangle$. We note that $\beta$ is time-independent, as well as $\lambda_{\pm}$, and thus can be computed at the start of the simulation. 
\par 
At this point, we can write $\textup{Diag}[\lambda_{+},\lambda_{-}]^{\otimes \lcg}$ as the direct sum of total spin states using the Clebsch-Gordan coefficients 
\begin{equation}
    \textup{Diag}[\lambda_{+},\lambda_{-}]^{\otimes \lcg} = \bigoplus_{j=0}^{\lcg/2} [M_j^{(0)}]^{\oplus B_j}\,,
\end{equation}
where we have defined
\begin{equation}
    M_j^{(0)} = \textup{Diag}\left[\lambda_{+}^{\lcg/2+j} \lambda_{-}^{\lcg/2-j}, ..., \lambda_{+}^{\lcg/2-j} \lambda_{-}^{\lcg/2+j}\right]\,,
\end{equation}
and
\begin{equation}
    B_j = \binom{\lcg}{\lcg/2-j} - \binom{\lcg}{\lcg/2-j-1}\,.
\end{equation}
The very same step can be performed for the second branch, just by swapping $\lambda_{+}$ and $\lambda_{-}$, obtaining $M_j^{(1)}$. In this representation, the basis of the density matrix $\tilde{\rho}_0(t)$ corresponds to $\{\vert j,m\rangle_{\vec{n}}\}$, while for the $\tilde{\rho}_1(t)$ corresponds to $\{\vert j,m\rangle_{\vec{n}'}\}$. In order to have a block diagonal representation of $\rho\Poilcg(t)$, we need to rotate back the two branch states in the same basis $\{\vert j,m\rangle_z\}$. To do so, we rotate each block with the rotation $R^j(-\gamma,\beta,\alpha)$, where $j$ corresponds to the representation of the corresponding block. In other words
\begin{equation}
    R^j(-\gamma,-\beta,-\alpha) = e^{i\gamma J_z}e^{i\beta J_y}e^{i\alpha J_z}\,,
\end{equation}
where now  $J_y$ and $J_z$ are the spin in the $j$th representation. In this way, we get into the new representation
\begin{equation}
    \tilde{\rho}_{x}(t)^{\otimes \lcg} = \bigoplus_{j=0}^{\lcg/2} [N^{(x)}_j(t)]^{\oplus B_j}\,,
    \label{eq:block-diag}
\end{equation}
with $x=0,1$ and where
\begin{equation}
    N_j^{(0)}(t) =  e^{i\gamma J_z}e^{i\beta J_y}e^{i\alpha J_z} M_j^{(0)} e^{-i\gamma J_z}e^{-i\beta J_y}e^{-i\alpha J_z}\,,
\end{equation}
and similarly for $N_j^{(1)}(t)$. In this way, both branch states are written on the same basis and the total state can be written as
\begin{equation}
    \rho\Pilcg(t) = \bigoplus_{j=0}^{\lcg/2} \left[p_0 N^{(0)}_j(t) + p_1 N^{(1)}_{j}(t)\right]^{\oplus B_j}\, .
\end{equation}
Given that we want to optimally discrimiante between the two branches, we need to find a measurement $\{\Pi_0(t),\Pi_1(t)\}$ that optimally discriminate between $\tilde{\rho}_0(t)^{\otimes\lcg}$ and $\tilde{\rho}_1(t)^{\otimes \lcg}$. To do so, we use optimal discrimination theory \cite{bergou2010discrimination} and we identify the optimal measurement for discrimination as the one to evaluate bias and agreement as per the definition in Eq. \eqref{eq:bia_bjb} and Eq. \eqref{eq:agr_bjb} respectively.
\par 
For binary discrimination of two equiprobable hypotheses $\rho_0$ and $\rho_1$, the strategy to find the optimal measurement is based on the operator
\begin{equation}
    \Lambda = \frac{1}{2}\left(\rho_0 - \rho_1\right)\, .
\end{equation}
The optimal measurement can be found by diagonalising this operator $\Lambda$. If we denote $\{\lambda_k, \vert \phi_k\rangle\}_k$ the eigenvalues and eigenvectors, we can number them in such a way that
\begin{align}
    \lambda_k < 0 & \text{  for  }  1\leq k < k_0 \\
    \lambda_k > 0 & \text{  for  }  k_0 \leq k \leq D \\
    \lambda_k = 0 & \text{  for  }  D < k \leq D_\Lambda\,,
\end{align}
where $D_\Lambda$ is the dimension of $\Lambda$. Then the optimal measurement for the discrimination is given as 
\begin{align}
    \Pi_1 & = \sum_{k=1}^{k_0-1} \vert \phi_k\rangle \langle \phi_k \vert \\
    \Pi_0 & = \sum_{k=k_0}^{D_\Lambda} \vert \phi_k\rangle \langle \phi_k \vert \,.   
\end{align}
This means that in the one-shot scenario, if we perform measurement $\{\Pi_1,\Pi_0\}$ and we observe outcome $x=0,1$, we can infer that the state is $\rho_x$ with some probability of error given as $P_{\textup{e}} = 1/2(1-\tr{\vert \Lambda\vert})$.
\par 
In our case, the two hypotheses corresponds to $\tilde{\rho}_0(t)^{\otimes \lcg}$ and $\tilde{\rho}_1(t)^{\otimes \lcg}$, and to compute the optimal measurement to discriminate them, we need to evaluate $\Lambda$ and its eigenvalues and eigenvectors. Thanks to the block diagonal matrix representation, this can be computed efficiently as follows. Given the representation in the spin basis $\{\vert j,m\rangle\}$ is given as Eq. \eqref{eq:block-diag}, we have that
\begin{equation}
    \Lambda = \frac{1}{2}\bigoplus_{j=0}^{\lcg/2}\left[N_j^{(0)}(t) - N_j^{(1)}(t)\right]^{\oplus B_j}\,.
\end{equation}
We clearly see here that we do not need to diagonalise the full $\Lambda$, but only each block $N_j^{(0)}(t)-N_j^{(1)}(t)$, whose maximum dimension scales as $2\lcg +1 \times 2\lcg +1$, i.e., linearly with $\lcg$. To each block corresponds the pair of eigenvalues and eigenvectors $\{\lambda_k^{(j)}, \vert \lambda_k^{(j)}\rangle \}$, and if $\lambda_k^{(j)}<0$ then the eigenvector will correspond to outcome $x=1$, and to otucome $x=0$ if $\lambda_k^{(j)} \geq 0$. By denoting $D^0_j$ the set of $k_j$ for which $\lambda_{k_j}^{(j)}\geq 0$ and $D^{1}_j$ the set of $k_j$ for which $\lambda_{k_j}^{(j)} < 0$, we have that on a specific block $\Pi^{(0)}_j = \sum_{k_j \in D^0_j} \vert \lambda_{k_j}^{(j)} \rangle \langle \lambda_{k_j}^{(j)} \vert$  and $\Pi^{(1)}_j = \sum_{k_j \in D^1_j} \vert \lambda_{k_j}^{(j)} \rangle \langle \lambda_{k_j}^{(j)} \vert$. We eventually get the optimal discrimination measurement as
\begin{equation}
    \Pi^{(x)} = \bigoplus_{j=0}^{\lcg/2} \left[\Pi^{(x)}_j\right]^{\oplus B_j}, \quad x =0,1 \, .
\end{equation}
This means that if we perform this pair of measurements, we would obtain
\begin{align}
    p_t(x\vert \lcg) = \tr{\Pi^{(x)} \rho\Pilcg(t)} & = p_0 \tr{\Pi^{(x)} \tilde{\rho}_0(t)^{\otimes \lcg}} + p_1 \tr{\Pi^{(x)} \tilde{\rho}_1(t)^{\otimes\lcg}}  = \\
    & = p_0 p_t(x\vert \rho_0(t)^{\otimes \lcg}) + p_1 p_t(x\vert \rho_1(t)^{\otimes \lcg})\,,
\end{align}
where
\begin{align}
    p_t(x \vert \rho_y (t)^{\otimes \lcg}) & = \tr{\Pi^{(x)}\tilde{\rho}_y(t)^{\otimes \lcg}} = \\
    & = \sum_{j=0}^{\lcg/2} B_j \tr{\Pi^{(x)}_j N_j^{(y)}(t)} = \\
    & = \sum_{j=0}^{\lcg/2} B_j \sum_{k_j \in D^{x}_{j}} \langle \lambda_{k_j}^{(j)} \vert N_j^{(y)}(t) \vert \lambda_{k_j}^{(j)} \rangle \,.
\end{align}
Using the latter form, we can thus evaluate efficiently the probabilities for the bias and agreement by just diagonalising the blocks $N_j^{(0)}(t)-N_j^{(1)}(t)$ and computing the latter formula. The bias thus becomes
\begin{equation}
    \Bias{U_{\star}(t)} = \frac{1}{2} \left( \left\vert p_t(0\vert \lcg) -p_0\right\vert + \left\vert p_t(1\vert \lcg) -p_1\right\vert\right)\,,
\end{equation}
while the agreement is given as
\begin{equation}
    \Agr{U_{\star}(t)} = \sum_{x=0}^1 \tr{\left[\Pi^{(x)} \right]^{\otimes \frac{\NP}{\lcg}} \rho\PiN(t)} = p_0 p_t(0\vert\lcg)^{\frac{\NP}{\lcg}} + p_1 p_t(1\vert\lcg)^{\frac{\NP}{\lcg}}\,.
\end{equation}
We report these two quantities as a function of time $t$ in Fig. \ref{fig:pure-deph-agr-bias}.
\begin{figure}[t]
    \centering
    \includegraphics[width=1\linewidth]{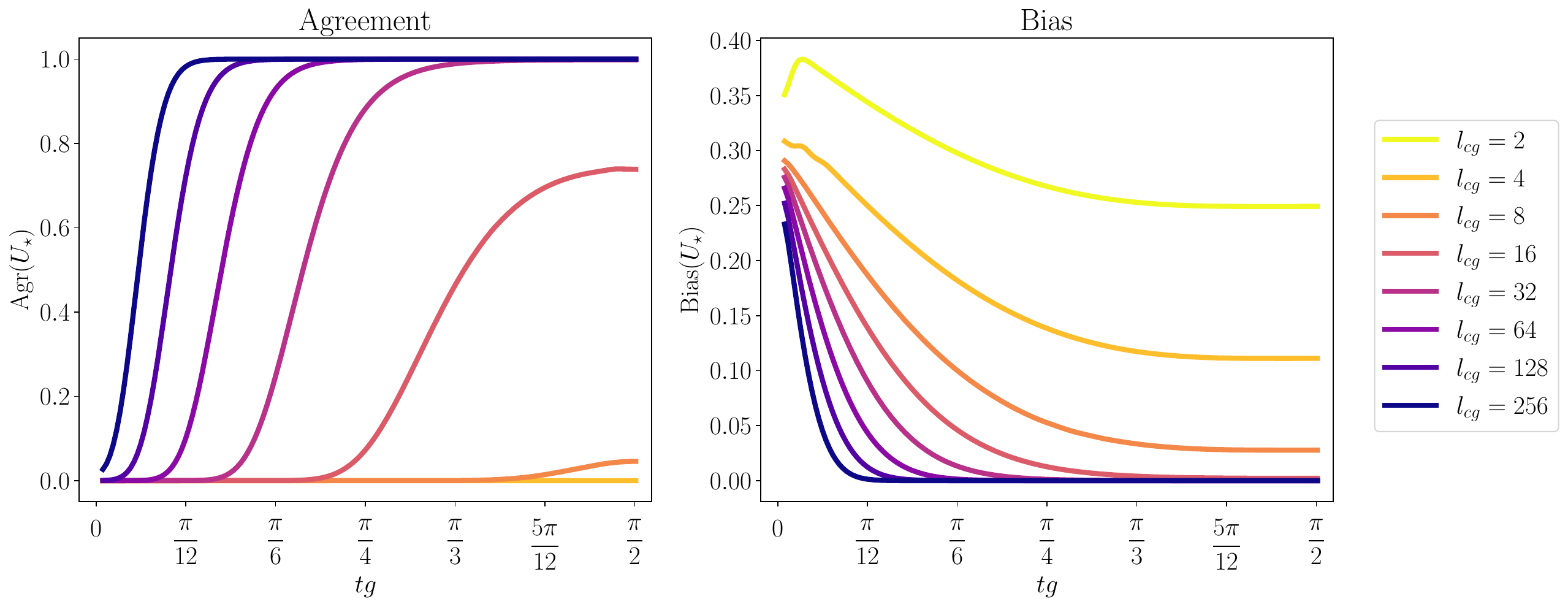}
    \caption{Plot of agreement and bias as a function of rescaled time $tg$, for an initial state of the system $\vert \psi_{0}\rangle = \sqrt{p_{0}} \vert 0 \rangle \langle + \sqrt{1-p_{0}} \vert 1 \rangle$ with $p_{0}=0.2$, while the environments are in a thermal state with inverse temperature $\beta=1$. Different colours correspond to different coarse-grainings $\lcg$. Here, $\NP=1024$ spins. The time scale $[0, \pi/(2g)]$, where the relevant dynamics occur, corresponds to half of the recurrence time $t_{\textup{r}} = \pi/g$.}
    \label{fig:pure-deph-agr-bias}
\end{figure}

\subsection*{Coarse-graining and time of measurement in the star spin model $H_{\star}$}
\label{app:cg-time}
In this subsection, we comment on the time evolution of agreement and bias for the $H_{\star}$ model, as reported in Fig.~\ref{fig:pure-deph-agr-bias}. For any tolerable error $\epsilon$ the agreement time $t_\epsilon$ is the minimum time for which $\text{Agr}(U_\star)\geq 1-\epsilon$.
\par
As we see from this Fig.~\ref{fig:pure-deph-agr-bias}, coarse-graining significantly influences the evolution of the two quantities in a similar manner. As $\lcg$ increases, both the maximum value and the duration for which we stay close to it increase. After some threshold value of $\lcg$, there is no visible difference in the maximum or minimum values. That difference becomes evident and can only be appreciated in the logarithmic plot shown in Fig.~\ref{fig:dis-bias-pure-deph}.
\par
However, even after reaching such a threshold, increasing $\lcg$ continues to have some effect, particularly on the threshold time at which the plateau is reached. Specifically, larger coarse-graining enables the two quantities to attain their ideal value more quickly and maintain it for a longer time, forming a stable plateau. This suggests that coarse-graining also impacts the speed of measurement, and we conjecture that there exists a threshold value of $\lcg$ after which the minimum time required to reach the stable plateau does not decrease as we increase $\lcg$. Investigating this behaviour in detail is left for future work.
\end{widetext}
\end{document}